\documentclass[aps,prd,amssymb,amsmath,eqsecnum,showpacs,nofootinbib,twocolumn]{revtex4-1}
\usepackage{color}
\usepackage{float}
\usepackage{graphicx}
\usepackage{ulem}
\usepackage{verbatim}
\usepackage{bigints}
\usepackage{hyperref}

\hypersetup{
    colorlinks=true,
    linkcolor=blue,
    citecolor=blue,
    urlcolor=blue
}

\usepackage{mathrsfs}

\begin{document}
\newcommand{\red }[1]{\textcolor{red}{#1}}
\newcommand{\ema}[1]{\textcolor{magenta}{#1}}
\newcommand{\blue}[1]{\textcolor{blue}{#1}}

\author{ Thomas M\"adler}
\email{thomas.maedler@mail.udp.cl}
\affiliation{Escuela de Obras Civiles and Instituto de Estudios Astrof\'isicos, Facultad de Ingenier\'{i}a y Ciencias, Universidad Diego Portales, Avenida Ej\'{e}rcito
Libertador 441, Casilla 298-V, Santiago, Chile. }

\author{Radouane Gannouji}
\email{radouane.gannouji@pucv.cl}
\affiliation{Instituto de Física, Pontificia Universidad Católica de Valparaíso\\ Av. Brasil 2950, Valparaíso, Chile}

\author{ Emanuel Gallo$^1$, $^2$}
\email{egallo@unc.edu.ar}
\affiliation{$^1$ Universidad Nacional de Córdoba, Facultad de Matemática, Astronomía, Física y Computación, Grupo de Relatividad y Gravitación; Córdoba, Argentina.\\$^2$ Consejo Nacional de Investigaciones Científicas y Técnicas,\\ CONICET, IFEG. Córdoba, Argentina. }

\title{Characteristic initial value problems for the Einstein-Maxwell-scalar field equations in spherical symmetry}

\begin{abstract}
The characteristic initial boundary problem is discussed in spherical symmetry for the Einstein-Maxwell-scalar field equations. It is formulated for an affine-null metric and the resulting field equations are cast into a hierarchical system of partial differential equations. The initial boundary value problem for a family of null hypersurfaces is specified for a timelike-null foliation at the central geodesic of spherical symmetry as well as for a double-null foliation where the corresponding boundary is a null hypersurface. For the latter, two distinct boundary value formulations arise; one where the null boundary has zero Misner-Sharp mass and another one where the corresponding Misner-Sharp mass is nonzero. As an application, the nonextremal and the extremal Reissner-Nordstr\"om solution in null coordinates for a charged black hole and the Fisher-Janis-Newman-Winicour solution are derived.

\end{abstract}

\maketitle
\section{Introduction}\label{sec:intro}

The choice of spacetime foliation with correspondingly adapted coordinates to solve  the Einstein field equations determines whether the resulting partial differential equations are either a complete set of hyperbolic partial differential equations or a mixed hyperbolic/elliptic system. 
The hypersurfaces of the hyperbolic part of this system along which the information travels are the characteristic surfaces.
Characteristic surfaces in general relativity may be for example  null cones along which gravitational waves propagate or null hypersurfaces  tied to the spacetime's causal structure. 
Solution methods for Einstein equations using the geometric information of the characteristic are commonly referred to as  characteristic methods. 
The employment of characteristic methods to study the field equations of general relativity very often brought a deeper understanding to various aspects of the theory.
Three important examples are the discovery of the Bondi mass loss formula  by Bondi, Sachs and collaborators \cite{Bondi:1960jsa, Bondi:1962px,Sachs:1962wk}, the first long term stable numerical evolutions of a black hole \cite{1998PhRvL..80.3915G} via computer simulations and recently the proof of incorrectness  of the third law of black hole thermodynamics \cite{Kehle:2022uvc}. 
The Bondi mass loss formula was found in an asymptotic patch of a spacetime where the radial coordinate parametrizes the null rays of outgoing null cones.
The long term stable black hole simulations were possible by numerically representing  the spacetime manifold with  a countable number of finite size patches discretised by finite difference operators adapted to either in- or outgoing null hypersurfaces.
The technical base of the disproof  of the third law was  the novel technique called ``characteristic gluing'' \cite{Aretakis:2021lzi}, where different characteristic (null) hypersurfaces are stitched together using gauge invariant quantities. 
These three examples have in common that the patch(es) is (are) equipped with at least one null coordinate, i.e. a coordinate constant along a null hypersurface, and that  boundary information at one side of the patch needs to be provided to achieve the physical result.  
For the Bondi mass loss case, this boundary data are the specification of an asymptotic inertial observer, an initial value for the mass and the so-called news function which is the retarded time derivative of the gravitational shear. 
For the black hole simulation, the boundary data were the specifications of black hole data at an inner world tube excising the black hole singularity, where the first patch of the numerically discretised null hypersurface is located. 
While in the disproof of the third law, the boundary data between the patches are  specifications of  quantities in Minkowski space and the horizon of a Schwarzschild/Reissner-Nordstr\"om black hole. 
The difference for the latter with respect to the former two is that the employed initial boundary value problem is based on the two intersecting null hypersurfaces instead of only one null hypersurface. 
A further difference is that in the last example the Einstein equations are solved  with a coupling to a charged scalar field, while the first two concerned the full Einstein equations in vacuum. 
 In spherical symmetry, the coupling to the scalar field captures  dynamical features and adds the necessary degrees of freedom so that the discussion of a characteristic initial boundary value problem (CIBVP) can be made without losing too much structural information. 

Here,  we investigate characteristic methods for Einstein equations minimally coupled to a  charged, massless and necessary complex scalar field in spherical symmetry. 
 In contrast to the numerous works \cite{Angelopoulos:2024yev,Kehle:2024vyt,Kehle:2023eni,Kehle:2022uvc,Kehle:2021jsp} related to characteristic gluing \cite{Aretakis:2021lzi} as well as the studies of formation of subextremal Reissner-Nordstr\"om black holes \cite{Murata:2013daa,Gelles:2025gxi} with characteristic methods using  a double-null foliation, we will investigate the CIBVP in affine-null metric coordinates \cite{Win2013}. 
The affine-null metric formulation is built on a foliation with respect to a family of null hypersurfaces and has the advantage that a simple evolution algorithm can be found, which is beneficial for numerical work. 
As we will demonstrate,  for this evolution scheme, the field equations can be cast into one set of  hierarchical hypersurface equations and one first order transport equation  with respect to the chosen null coordinate. The set of derived equations is similar to a previously used one for a real scalar field \cite{Crespo:2019mcv} having the  additional advantage that it can be used to investigate black hole spacetimes beyond the formation of an event horizon and track the apparent horizon and the scalar field inside the black hole \cite{Madler:2024kks}. 
This  interior   black hole evolution is possible because the hypersurface-evolution equations remain regular where the   expansion rate of the outgoing null rays vanishes, i.e. at an apparent horizon. 

 {In fact, with an extension of \cite{Crespo:2019mcv},  it is possible to investigate the Choptuik phenomenon \cite{Choptuik:1992jv} for charged scalar fields by globally foliating spacetime to include null infinity on the numerical grid through coordinate compactification}\footnote{ {It was shown in \cite{Madler:2025oiy} that the conformal Einstein-scalar field equations in affine-null coordinates \cite{Friedrich:1986rb,Penrose:1962ij} are equivalent to corresponding physical Einstein-scalar field equations, provided the latter employ a compactified affine parameter and suitable normalizations of the metric and scalar field. Thus, null infinity can be represented on a numerical grid using a compactified coordinate.}}. 
 {Traditionally, such investigations have been carried out only within the 3+1 formulation of the Einstein-Maxwell-scalar field equations \cite{Gundlach:1996vv}, which does not usually allow this.
These studies can also address the formation process of charged, spherically symmetric black holes.
Indeed, the CIVBP presented below can be used to numerically test the mathematical results of \cite{Kehle:2018zws}, particularly regarding the disproof of the third law of black hole thermodynamics, that is, as a consequence of \cite{Kehle:2022uvc}, extremal black holes should form in finite time.}  {
 In a numerical simulation, this would imply identifying an extremal black hole in the output of a simulation where the black hole forms within finite simulation time.
In this context, it is significantly simpler to study the formation of extremal black holes in the Einstein-Maxwell-scalar field system in spherical symmetry rather  than to study the formation of extremal Kerr black holes.
This is because, in the former, the simulations can be performed with a single spherically symmetric dynamical field (i.e., the scalar field)} {, whereas the latter requires at least axial symmetry for the Einstein equations, involving two dynamical fields (i.e. the gravitational wave degrees of freedom).}

 {Affine-null coordinates can also be useful for constructing outgoing null coordinate systems that not only describe the exterior of black holes, but are also sufficiently regular to describe the region inside the event horizon. 
 Coordinate systems of this kind for the Schwarzschild case include, for example, the Israel coordinates \cite{Israel:1966zz}, which were later generalized to include a broader family of metrics in\cite{Gallo:2021jxt}. Other null coordinate systems have been constructed numerically for the Kerr black hole in \cite{Arganaraz:2021fpm}, and used to describe the evolution of a scalar field penetrating the black hole interior in \cite{Arganaraz:2022mks}. 
 A fully analytical construction of an affine-null coordinate system to describe slowly rotating Kerr black holes is also possible, as demonstrated in \cite{Madler:2023aqc}. The general formalism presented in this work for spherically symmetric spacetimes accommodates such regular affine-null coordinate systems as a particular case, thereby allowing for the study of matter fields radiating in the exterior region while simultaneously tracking their evolution as they penetrate into the interior of black holes.}

 {In this work, we extend and modify the scheme of \cite{Crespo:2019mcv} so that the regularization of the fields involved in the hierarchical affine-null system of equations is preserved when the scalar field is complex and a Maxwell field is included in the energy-momentum tensor.}
We will also show that for the derived set of equations,  CIBVPs can be formulated either on the central world line in spherical symmetry or  by two different versions of  double-null CIBVPs at a null boundary. 
As an example for these cases, we discuss the asymptotic CIBVP showing that the entire asymptotic behavior of a charged scalar field is governed by the specification of an initial value for the scalar field monopole, mass, charge and a free function -  the scalar field news function.
We further derive the subextremal and extremal Reissner-Nordstr\"om black hole as well as the Fisher-Janis-Newman-Winicour (FJNW) solution in the  affine-null-metric coordinate chart. 
Those solutions turn out to be expressible by simple rational functions.

In Sec.~\ref{sec:fieldEQ}, the basic field equations and additional geometric quantities for the Einstein-Maxwell-scalar field equations in spherical symmetry and affine-null metric coordinates are presented.
The regularized set of hypersurface-evolution equations following from the field equations being the  CIBVPs is given in Sec.~\ref{sec:cibvp}. 
The  CIBVP in an asymptotic region is presented in Sec. \ref{sec:asympt_cibvp}, the one at the central geodesic is formulated in \ref{sec:cibvp_worldline}, while the two different CIBVPs with a null boundary surface are given in Sec.~\ref{sec:cibvp_dn}. 
In Sec.~\ref{sec:symmetry}, we discuss the properties for asymptotic timelike Killing vectors, which are needed for the derivation of the Reissner-Nordstr\"om black hole in Sec.~\ref{sec:RN_null} and the FJNW solution in Sec.~\ref{sec:FJNW_null}. Section \ref{sec:summary} contains a summary and conclusion of the work.

We use geometric units with  $c=1$ and the Einstein gravitational constant is $\kappa=8\pi G$. The metric signature is $+2$, capital Latin indices $A,B,C,\hdots$ are two dimensional with values $x^A = (x^2, x^3)$ and  lowercase Latin indices are four dimensional taking values $x^a = (w,\lambda, x^A)$. The covariant derivative of the metric is $\nabla_a$ and  partial derivatives are denoted with the comma notation $\frac{\partial f}{\partial x^a} = f_{,a}$.
Calligraphic capital letters, $\mathcal{J},\mathcal{L},\mathcal{C}, ...$ are always assumed complex fields, while Roman capital letters, such as  $Q, Z, V, L,\hdots$ are purely real.
Numeral subscripts in parentheses, like $V_{(1)}$, correspond to expansion coefficients of a  power series evaluated at $\lambda=0$ with   the number in the parentheses referring to the power $\lambda^k$. Numeral subscripts in brackets, such as $\Phi_{[1]}$, correspond to expansion coefficients of an  asymptotic  power series and the number in the bracket refers to the order of the inverse power $1/\lambda^k$.

\section{Field equations  for the Einstein-Maxwell-scalar field equations}\label{sec:fieldEQ}

In a four-dimensional spacetime, consider the   Einstein-Hilbert action $S$ given by
\begin{equation}
S=
 \int d^4x \sqrt{-g}\left[\frac{ g^{ab}R_{ab} }{2\kappa} 
 -\frac{ F_{ab}F^{ab}}{16\pi}
- \frac{1}{2}(\overline{\mathcal{D}^a\Phi})(\mathcal{D}_a\Phi) 
\right],
\end{equation}
where $\kappa = 8\pi G$ is the Einsteinian gravitational  constant, $\sqrt{-g}$ is the determinant of the metric $g_{ab}$ with a well defined inverse $g^{ab}$ and a corresponding Levi-Civita covariant derivative $\nabla_c$, $R_{ab}$ is the Ricci tensor with respect to $g_{ab}$, $F_{ab} = \nabla_a A_b - \nabla_b A_a$ is a Maxwell field determined by the four-covector $A_b$, and $\mathcal{D}_c = \nabla_c + iqA_c$ is a gauge invariant derivative for the massless complex scalar field $\Phi$ that is coupled to the Maxwell field with the gauge charge $q$. The overbar denotes complex conjugation. 
From the variation of the action with respect to   $g^{ab}$, $A_a$, $\bar \Phi$ and $\Phi$, we can derive the trace reversed Einstein equations, $E_{ab}$, the Maxwell equations, $E_a$, and the scalar wave equations $E$ and $\bar E$, which are, respectively 
\begin{eqnarray}
E_{ab}:= &0&=R_{ab} -  \frac{\kappa}{4\pi}\left( g^{cd}F_{ac}F_{bd}
-\frac{1}{4}  g_{ab} F^{cd}F_{cd}\right)
\nonumber\\
&& 
-\kappa\left[ \frac{1}{2}(\overline{\mathcal{D}_a\Phi})(\mathcal{D}_b\Phi)
+ \frac{1}{2} (\overline{\mathcal{D}_b\Phi})(\mathcal{D}_a\Phi)
  \right], \\
E_a:=&0&=\nabla_a F^{a}_{\;b} + 4\pi j_b,\\
E:=&0&={\mathcal{D}^a}\mathcal{D}_a \Phi, 
\\
\bar E:=&0&=\overline{\mathcal{D}^a\mathcal{D}_a \Phi} ,
\end{eqnarray}
with 
the current 
\begin{equation}
j_b =
\frac{iq}{2} \left(\overline{\Phi}\mathcal{D}_a\Phi  -\Phi\overline{\mathcal{D}_a \Phi} 
 \right),
\end{equation}
whose integral over a three-volume $\mathtt{V}$ gives  rise to the charge function\footnote{Stokes theorem for an antisymmetric tensor $F^{ab}$ is \cite{Poisson:2009pwt}
$$\int_\mathtt{V} F^{ab}_{\;\;\;;b} d\Sigma_a= \frac{1}{2}\oint_{\partial \mathtt{V}}F^{ab} d\Sigma_{ab},$$ where $\mathtt{V}$ is a three-dimensional hypersurface with the volume element $d\Sigma$  and where $\Sigma =\partial\mathtt{V}$ is  its  two-dimensional boundary having the surface element $d\Sigma_{ab}$. For a two surface $\partial \Sigma$ with metric $h_{AB}$ with $h=\det(h_{AB})$ and equipped with coordinates $h^A = (h^1, h^2)$, $d\Sigma_{ab} = 2\ell_{[a}n_{b]}\sqrt{h}dh^1dh^2$ where $\ell^{a}$ and $n^a$ are two null vectors obeying $n_a\ell^a = -1$.}
\begin{equation}\label{eq:chargeFunction}
  Q 
= \int_\mathtt{V} j^ad\Sigma_a 
=\frac{1}{4\pi}\oint_{  \Sigma}  F^{ab}\ell_{[a}n_{b]} \sqrt{\det(g_{AB})}dx^2dx^3 , 
\end{equation}
with $g_{AB}$ being the two-metric of the spacelike hypersurface of the co-dimension two boundary  $\Sigma=\partial\mathtt{V}$, which is the intersection of the integral curves of null vectors $\ell^a\partial_a$ and $n^a\partial_a$  orthogonal to $\Sigma$. 
Consider  the metric 
\begin{equation}\label{eq:metric}
ds^2 = -V(w,\lambda)dw^2+2\epsilon dw d\lambda + r^2(w , \lambda)q_{AB}dx^Adx^B,
\end{equation}
where $q_{AB}$ is any metric of a unit sphere\footnote{In standard spherical coordinates $x^A=(\theta,\varphi)$, ${q_{AB} = \mathrm{diag}(1, \sin^2\theta)}$ with the three associated Killing vectors $\xi_{1} = \partial_\phi$, $\xi_2 =  \cos\phi  \partial_\theta - \cot\theta \sin\phi  \partial_\phi$ and $\xi_3 = \sin\phi  \partial_\theta + \cot\theta \cos\phi  \partial_ \phi$. In principle, this two metric could also be a flat metric, like ${q_{AB} = \mathrm{diag}(1,1)}$, or a unit hypersphere ${q_{AB} = \mathrm{diag}(1, \sinh^2\theta)}$. Of the field equations, only \eqref{eq:qABEAB} would change by the substitution of the term +1 by either $+0$ or $-1$, respectively. The former may be interesting for plane wave front propagation, such as pp waves, while the latter for hyperbolical black holes or fluids (see e.g. \cite{Herrera:2020bfy,Herrera:2021pbq}).}. 
An  understanding of the binary parameter $\epsilon$ can be found in flat space, where $V=1$ and $r=\lambda$, then for  $\epsilon=-1$ the coordinate surfaces $w=$const represent  outgoing null hypersurfaces $u$=const, while for $\epsilon=1$ they are ingoing null hypersurfaces $v=$const. 

The inverse metric has the components
\begin{equation}\label{eq:inverse_g}
g^{w\lambda} = \epsilon\;\;,\;\;g^{\lambda\lambda} = V\;\;,\;\; g^{AB} = \frac{1}{r^2}q^{AB}\;,
\end{equation}
where we used $\epsilon^2=1$. 
The form of the metric \eqref{eq:metric} remains invariant under the coordinate changes
\begin{equation}\label{eq:invariance_metric}
    w=\int^w a(\hat w)d\hat w\;\;,\;\;\lambda = a(\hat w)\hat \lambda + b(\hat w).
\end{equation}
For the metric, we define two  null vectors
\begin{equation}
\ell^a\partial_a = \partial_\lambda\;\;,\;\;
n^a\partial_a = -\epsilon\partial_w -\frac{V}{2}\partial_\lambda\;,
\end{equation}
whose associated one forms are $\ell _a dx^a = \epsilon dw$ and $n_adx^a = \epsilon (V/2)dw -d\lambda$ satisfying $n^a\ell_a = -1$. 
The null vector field $\ell^a\partial_a$ is tangent to affinely parametrized null geodesics meaning $\ell^a\nabla_a \ell^a = 0$,
while the  vector field $n^a\partial_a$ is in general tangent to nonaffinely parametrized null geodesics, where
\begin{equation}\label{eq:n_inaffine}
    n^a\nabla_a n^b = -\frac{1}{2}V_{,\lambda}n^a\;\;.
\end{equation}

The electromagnetic field strength tensor $F_{ab}$ assumed to be spherically symmetric, is invariant under the gauge transformation $A_\mu\rightarrow A_\mu+\chi_{,\mu}(x^a)$. 
As of spherical symmetry, the gauge field $\chi$ depends only on $w$ and $\lambda$. 
Choosing $\chi = -\int_{\lambda_0}^\lambda A_\lambda d\lambda$ yields  the null gauge condition $A_\lambda=0$. 
The remaining gauge freedom allows us to set $\lim _{\lambda\rightarrow\lambda_\Gamma}A_w(w,\lambda)=0$ where the limit toward $\lambda_\Gamma$ is taken toward a world line, a finite size world tube, or a null hypersurface (such as a horizon or null infinity). 
The four-covector  has the simple form  $A_adx^a = \alpha dw$ which implies  $A^a\partial_a = \epsilon \alpha \partial_\lambda$ and $A_bA^b = 0$. 
The Maxwell fields are $F_{w\lambda}=-F^{w\lambda}=-\alpha_{,\lambda}$, $F^w_{\;w} = -F^\lambda_{\;\lambda} = \epsilon\alpha_{,\lambda}$ and $F_{ab}F^{ab} = -2(\alpha_{,\lambda})^2$. 
For the charge function \eqref{eq:chargeFunction}, we find
\begin{align}
Q =  -\epsilon r^2\alpha_{,\lambda}.
\end{align}
The nonzero components of  $E_{ab}$ are 
\begin{widetext}
\begin{eqnarray}
\label{eq:Eww}
(E_{ww})&:&0=
 r_{,ww}   
-\frac{V (r^2V_{,\lambda})_{,\lambda} }{4r}
-\frac{\epsilon}{2}\left(
 r_{,w}V_{,\lambda}  - r_{,\lambda}V_{,w} \right)
+\frac{1}{2}\kappa r\left[
|\Phi_{,w}|^2 
+  \frac{VQ^2}{8\pi r^4}  
+ q^2 \alpha^2 |\Phi|^2
 + iq  \alpha \Bigl(\Phi\bar{\Phi}_{,w}-\bar{\Phi}\Phi_{,w}\Bigr)
   \right]\;,\label{eq:E00}
\nonumber\\%
&&\\
(E_{w\lambda})\;&:&
0=r_{,w\lambda} 
  + \frac{\epsilon}{4r}(r^2V_{,\lambda})_{,\lambda} 
  - \frac{1}{2}\kappa r\left[
   \frac{\epsilon Q^2  }{8\pi r^4}  
-\frac{1}{2}\left( \overline{\Phi}_{,w} \Phi_{,\lambda}
+\overline{\Phi}_{,\lambda} \Phi_{,w} \right)
- \frac{iq\alpha}{2}(\Phi  \overline{\Phi}_{,\lambda} -\overline{\Phi}  \Phi_{,\lambda})  
\right]\label{eq:E01}\\
(E_{\lambda\lambda})\;
&:&0=
 r_{,\lambda\lambda}
 +\frac{1}{2}\kappa r |\Phi_{,\lambda}|^2\;,\label{eq:E11}
 \\
(q^{AB}E_{AB})\;
&:&0=
- \left( Vrr_{,\lambda}  +  2\epsilon rr_{,w} -\lambda\right)_{,\lambda} 
-\frac{\kappa }{8\pi }\frac{Q^2}{r^2},
 \label{eq:qABEAB}
\end{eqnarray}
and those of $E_a$ are
\begin{eqnarray}
 (E_w)\;&:&0= Q_{,w}+\epsilon V Q_{,\lambda}+4\pi q^2 r^2 \alpha |\Phi|^2
 +2i\pi q r^2\Bigl(\Phi\bar{\Phi}_{,w}-\bar{\Phi}\Phi_{,w}\Bigr)\label{eq:E0}\;,\\
(E_\lambda)\;&:&0=  Q_{,\lambda} 
+2\pi  r^2 q i(\overline {\Phi} \Phi_{,\lambda} - \Phi   \overline{ \Phi}_{,\lambda})\label{eq:E1}\;,
\end{eqnarray}
while $E$ gives
\begin{equation}
0=
(r^2\Phi_{,w})_{,\lambda}
+ (r^2\Phi_{,\lambda})_{,w}
+\epsilon (r^2V\Phi_{,\lambda})_{,\lambda}
+2iqr  \alpha (r\Phi) _{,\lambda} 
-i\epsilon  q\Phi Q \;.
\label{eq:E}
\end{equation}
\end{widetext}

In spherical symmetry the Misner-Sharp mass 
 \cite{Misner:1964je},
\begin{align}
\label{eq:misner_sharp_mass}
    M = \frac{r}{2}(1-g^{ab}r_{,a}r_{b})
    =  \frac{r}{2}\Bigl(1-2\epsilon r_{,w}r_{,\lambda}-Vr_{,\lambda}^2\Bigr),
\end{align}
is an invariant concept of quasilocal mass.  Assuming asymptotically flat spacetimes,  its asymptotic limit,
\begin{equation}\label{eq:Bondi_mass}
    m_B:=\lim_{\lambda\rightarrow\infty}M,
\end{equation}
is the Bondi mass $m_B$. 
With the complex null vector $m^a\partial_a = r^{-1}q^A\partial_A$ in which the complex dyad $q^A\partial_A$ is any dyad for the unit sphere metric normalized like  ${q^Aq_A = q^A\bar q_A-1=0}$, we calculate  the Weyl scalar ${\Psi_2= C_{abcd}\ell^am^b\bar m^ c n^d}$  \cite{Griffiths:2009dfa} like
\begin{align}
\Psi_2 
=&
-\frac{1}{6r^2}
-\frac{\epsilon}{3}\left(\frac{r_{,\lambda}}{r}\right)_{,w}
-\frac{ V}{6}\left(\frac{r_{,\lambda}}
{r}\right)_{,\lambda}
+\frac{r^2}{12}\left(\frac{V_{,\lambda}}
{r^2}\right)_{,\lambda}\;,
\end{align}
whose asymptotic limit provides an alternative way to calculate the Bondi mass  in a frame approaching  an inertial frame at large values for the affine parameter 
\begin{equation}
m_B = -\lim_{\lambda\rightarrow\infty}\left[\lambda^ 3\Psi_2 - \frac{1}{12}\kappa \epsilon \lambda^2 \left(\Phi\bar{\Phi}_{,w}+\Phi_{,w}\bar{\Phi}\right)\right]\;.
\end{equation}

\section{The characterisitic initial-boundary value problem}\label{sec:cibvp}
The formulation of the CIBVP for \eqref{eq:E00} - \eqref{eq:E} is a consequence of  the twice contracted Bianchi identities, $\nabla^a(E_{ab} - \frac{1}{2}g_{ab}(g^ {cd}E_{cd}))$ and its electromagnetic counterpart $\nabla^a E_a$ and it follows the idea of \cite{Bondi:1962px,Sachs:1962wk, Sachs:1962zzb,Tamburino:1966zz}. 
The spirit\footnote{ {See e.g. \cite{Sachs:1962zzb,Tamburino:1966zz} or Appendix A of  \cite{Madler:2018bmu} for a discussion on the twice contracted Bianchi identity for a general metric adopted to a family of null hypersurfaces and the classification of main equations, supplementary equations and trivial equation.}} is to pick from $(E_{ab}, E_a, E)$ a set of main equations to hold everywhere, and then the  twice contracted Bianchi identities and this set of main equations imply that the complementary equations of  the main equations either hold trivially as a consequence of the main equations or they must be satisfied at a given value of $\lambda=\lambda_{\mathcal{B}}$=const. 
The former is called the trivial equation and the latter are the so-called supplementary equations. For the set \eqref{eq:E00} - \eqref{eq:E},  the following assignations are made: 
\begin{itemize}
    \item {\bf  main equations         } 
    \begin{equation}\label{eq:main}
        E_{\lambda\lambda}\;,\;\;q^{AB}E_{AB}\;,\;\;E_\lambda\;,\;\;E\;,
    \end{equation}
    which are given by \eqref{eq:E11}, \eqref{eq:qABEAB}, \eqref{eq:E1}, and \eqref{eq:E}, respectively.
    \item {\bf supplementary equations : } 
    \begin{equation}
        E_{ww}\;,\;\;E_w,
    \end{equation}
    given by \eqref{eq:E00} and \eqref{eq:E0}, respectively
    \item {\bf trivial equation      : } 
    \begin{equation}
        E_{w\lambda},
    \end{equation}
    represented  by \eqref{eq:E01}.
\end{itemize}

 {In the Bondi-Sachs formulation of the CIBVP \cite{Bondi:1962px,Sachs:1962wk,Madler:2016xju} the main equations form a natural hierarchical system of hypersurface equations that can be solved provided there is an initial profile of the dynamically propagating field(s). 
That is, each of the equations is solved one after another, in successive order.
 For the Einstein-scalar field equations in spherical symmetry while equipped with Bondi-Sachs coordinates, the only field propagating is the scalar field. There are two remaining fields in the system parametrizing the metric, the fields $\beta_{BS}$ and $V_{BS}$ (see, e.g., \cite{Gomez:1994ns,Madler:2025oiy}).
Indeed, in that case [where $E_\lambda$ in \eqref{eq:main} does not need to be considered], $E_{\lambda\lambda}$ is integrated first with the initial data to find $\beta_{BS}$. 
Then, with the initial data and $\beta_{BS}$, the field $V_{BS}$ is found by the integration of $q^{AB}E_{AB}$ in \eqref{eq:main}. 
Next, the wave equation $E$ in \eqref{eq:main} is used to update the initial data from the initial null hypersurface to a null hypersurface at later times.} 

 {However, this beautiful and beneficial hierarchy for numerical integration is destroyed in the affine-null formulation, due to the appearance of the $r_{,w}$ derivative in (\ref{eq:qABEAB}). 
This additional time derivative is a consequence of the affine-null coordinate choice. 
The affine parameter $\lambda$ is related to the Bondi-Sachs area distance coordinate $r$ via $\lambda_{,r}(w,r,x^A) = \exp[2\beta_{BS}(w,r,x^A)]$. 
The  transformation appears simple, but it actually comes along with a change in the $\partial_w$ directions}\footnote{ { {The $\partial_A$-directions change in addition, if  the assumption of spherical symmetry is dropped.}}}.  {This results in the additional $w$-derivative in \eqref{eq:qABEAB} \cite{Win2013}.
}

 {Nevertheless, we can establish the hierarchy in the affine-null case after the introduction of the following new fields}
\begin{subequations}\label{eq:def_hierarchy_vars}
\begin{eqnarray}
\rho&=&r_{,w}\;\;,\;\;\\
\label{eq:defZ}
Z &= & Vrr_{,\lambda}  + 2\epsilon r\rho - \lambda\;\;,\;\;\\
\label{eq:defL}
 \mathcal{L} &=& \frac{2r\left(  r_{,\lambda}\Phi_{,w}-\rho\Phi_{,\lambda}\right)+\epsilon (Z+\lambda)\Phi_{,\lambda}}{r_{,\lambda}}.
\end{eqnarray}
\end{subequations}
which cast \eqref{eq:main} into
\begin{subequations}\label{hyp_ev}
\begin{eqnarray}
0
&=& 
 r_{,\lambda\lambda}
 +\frac{1}{2}\kappa r |\Phi_{,\lambda}|^2\label{eq:hyp_r}
 \;,\\
 Q_{,\lambda} &=& 
-2\pi  r^2 q \mathcal{J}\;,\\
\label{eq:hyp_a}
\alpha_{,\lambda}&=&-\epsilon \frac{Q}{r^2}\;,\\
Z_{,\lambda} 
&=&
-\frac{\kappa }{8\pi }\frac{Q^2}{r^2}
\label{eq:hyp_Z}\;,
 \\
\mathcal{L}_{,\lambda}&=&
      -\frac{\epsilon (\lambda + Z)\Phi_{,\lambda}}{r } 
   - 2iq  \alpha (r\Phi) _{,\lambda} 
+i\frac{\epsilon  q\Phi Q}{r} \label{eq:hyp_L}
 \;,  \\
V_{,\lambda\lambda} &=&
 -\frac{1}{\lambda}\left(\frac{\lambda^ 2 }{r^2 }\right)_{,\lambda}
  +\frac{ 2 Z r_{,\lambda}}{r^3}
   +\frac{\kappa}{2\pi}\frac{ Q^2}{r^4}   
   \nonumber\\
   &&
   - \frac{\kappa \epsilon}{2r }\left(\overline{\Phi_{,\lambda}} \mathcal{L} + \overline{\mathcal{L}}\Phi_{,\lambda} \right)
  +  \kappa q \epsilon\alpha    \mathcal{J} 
 \;, \label{eq:hyp_V}
\\
\Phi_{,w} &=& \frac{\mathcal{L}}{ 2r } - \frac{\epsilon}{2}V\Phi_{,\lambda}\;,\label{ev_eqn_Phi}
\end{eqnarray}
where we introduced the auxiliary variable
\begin{equation}\label{eq:def_J}
\mathcal{J} = i(\overline {\Phi} \Phi_{,\lambda} -  \overline{ \Phi}_{,\lambda}\Phi  ).
\end{equation}
\end{subequations}

This system is completed with the two supplementary equations \eqref{eq:E00} and \eqref{eq:E0} as well as \eqref{eq:qABEAB} 
evaluated at the boundary $\lambda=\lambda_\mathcal{B}$ = const, which can be a world line \cite{Madler:2012sg}, a finite size world tube \cite{Winicour:2011jn} or a null hypersurface \cite{Madler:2018bmu}. 

A solution of  \eqref{hyp_ev} requires initial data 
\begin{equation}
    \Phi(0, \lambda) = \mathcal{F}_{\Phi}(\lambda),
\end{equation}
on a null hypersurface $w=0$ for values $\lambda\ge\lambda_{\mathcal{B}}$, where $\mathcal{F}_{\Phi}(\lambda)$ is an arbitrary function. 
It further requires the following boundary conditions at $\lambda=\lambda_\mathcal{B}$  for all values $w\neq 0$
\begin{equation}
    \begin{split}
        &r(w, \lambda_\mathcal{B})\;,\;\;
r_{,\lambda}(w, \lambda_\mathcal{B})\;,\;\;
Q(w, \lambda_\mathcal{B})\;,\;\;
\alpha(w, \lambda_\mathcal{B})\;,\;\;\\
&
\mathcal{L}(w, \lambda_\mathcal{B})\;,\;\;
V(w, \lambda_\mathcal{B})\;,\;\;
V_{,\lambda}(w, \lambda_\mathcal{B})\;.
    \end{split}
\end{equation}

In particular, the boundary conditions  for the variables $r$, $Q$, $\alpha$, and $V$ follow from physical considerations, while those for auxiliary  fields $Z$ and $\mathcal{L}$  follow from the definitions \eqref{eq:def_hierarchy_vars} and are given by
\begin{subequations}    
\label{eq:bc_LY}
\begin{eqnarray}
    Z|_{\lambda=\lambda_\mathcal{B}} &=& (Vrr_{,\lambda} + 2\epsilon rr_{,w}-\lambda)|_{\lambda=\lambda_\mathcal{B}} \;\;,\;\;\label{eq:bc_Z}\\
    \mathcal{L}|_{\lambda=\lambda_\mathcal{B}} &=& 
    \left(2r\Phi_{,w} + \epsilon r V\Phi_{,\lambda}\right)|_{\lambda=\lambda_\mathcal{B}}\;\;.
\end{eqnarray}
\end{subequations}
The $r_{,w}$ derivative in the boundary condition for $Z$ in \eqref{eq:bc_Z} can be either removed or evaluated from  $\Phi_{,w}$  via a differential equation  in the formulation of the CIBVP (see next sections).
Thus, the boundary conditions depend only on the  $w$-derivative of the propagating physical field i.e.,  the  derivative $\Phi_{,w}$.
Consequently, the only propagating degree of freedom of the formulated CIBVP is the field that is subject to a hyperbolic equation, whose characteristics are given by the $w=$const null hypersurfaces. 
For the Einstein-Maxwell-scalar-field equations in spherical symmetry only  the scalar field is determined by a hyperbolic partial differential equation, i.e., the wave equation $\mathcal{D}_a\mathcal{D}^a \Phi=0.$

Since we can always translate the  origin of an affine parameter by a transformation like $\lambda\rightarrow a(w)\lambda+b(w)$, we set hereafter $\lambda_\mathcal{B}=0$. 
In the following three sections we discuss the asymptotic initial value problem, the initial value problem at a world line, and the double-null initial boundary value problem for two intersecting null hypersurfaces.


\subsection{Asymptotic initial value problem for large values of the affine parameter}\label{sec:asympt_cibvp}

\begin{figure*}[htbp]
  \centering
  \includegraphics[width=0.6\linewidth]{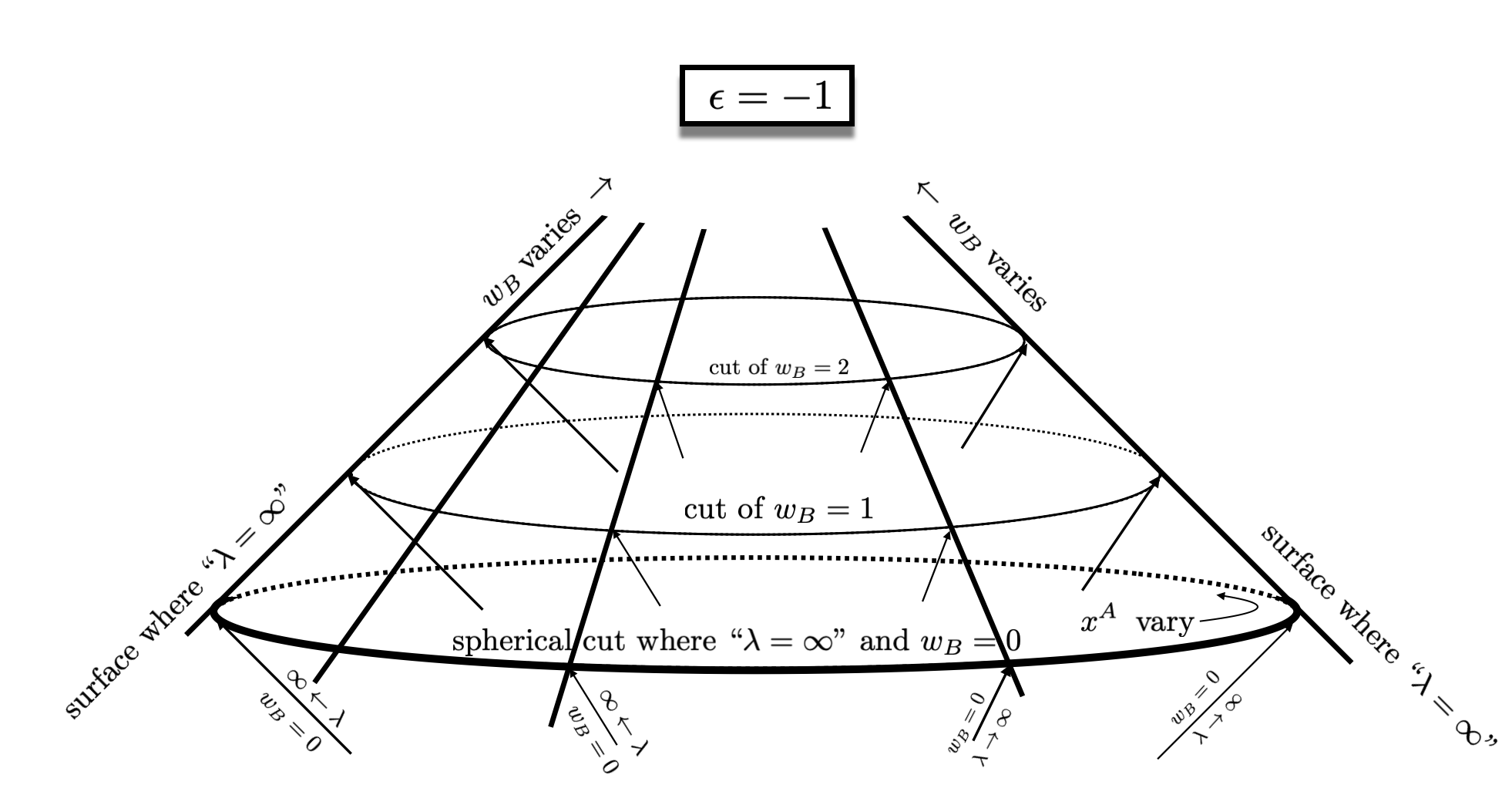}
  \includegraphics[width=0.5\linewidth]{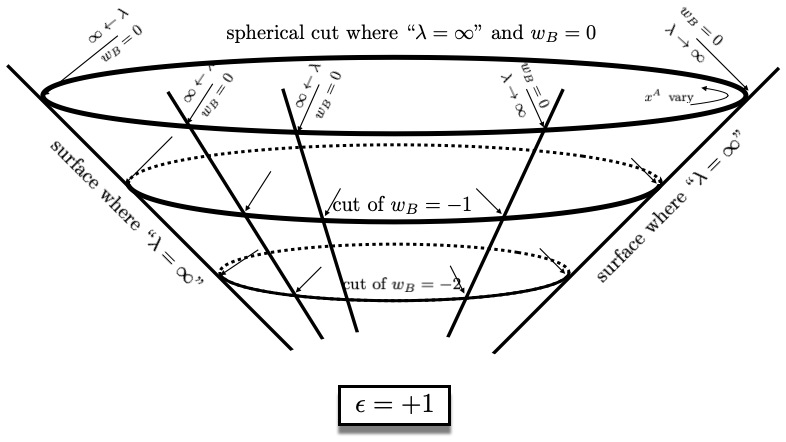}
  \caption{
   {
  The asymptotic null hypersurface for the cases $\epsilon=\pm1$ are shown by spherical cuts at the location of ``$\lambda=\infty$''.
  Either the idealized limit ``$\lambda=\infty$'' can be attached to the physical spacetime via the conformal Penrose compactification \cite{Penrose:1962ij} or it can  be achieved by introducing a radial compactified coordinate, such as $\ell=1/\lambda$, and a proper regularization of the field equation\cite{Madler:2025oiy}.
  The upper graphic illustrates a series of spherical cross sections of a Bondi frame, where the affine parameter $\lambda\rightarrow\infty $ on outgoing null hypersurfaces $w_B$=const. The Bondi time $w_B$ then corresponds to the retarded time $u_B$. 
  The lower graphic shows the corresponding situation for ingoing null hypersurfaces toward a Bondi frame, where $\epsilon=+1$ and $w_B$ are ingoing null hypersurfaces. In this case, the Bondi time is the advanced coordinate $v_B$. In both graphics, coordinates $x^A$ parametrize the cuts at $w_B$=const and label the different rays (indicated by the arrow lines  also carrying $\lambda\rightarrow\infty$ ) reaching the ``$\lambda=\infty$'' (null) hypersurface  . }
   }
  \label{fig:infinities}
\end{figure*}

 {To consider the asymptotic characteristic initial value problem, we assume that the fields have sufficient differentiability for large values of the affine parameter and in the limit $\lambda\rightarrow \infty$.  An illustration of the null hypersurfaces for the parameter $\epsilon$ is given in Fig.~\ref{fig:infinities}. 
The scalar field is assumed to obey the radiation condition that the limit $\lim_{\lambda\rightarrow\infty}\lambda\Phi$ is finite on $w=$const null hypersurfaces.} 
Then the scalar field has the following asymptotic expansion
\begin{equation}\label{eq:phi_inf}
    \Phi  =
    \frac{\varphi_{[1]} +i\psi_{[1]} }{\lambda} 
    +
     \frac{\varphi_{[2]} +i\psi_{[2]} }{\lambda^2}
     + O(\lambda^ {-3})\;,
\end{equation}
where the functions with subscript indices in brackets labeling the order of $\lambda^{-i}$ are real functions depending on $w$. The function $\mathcal{L}$ is also decomposed into real and imaginary parts
\begin{equation}
    \mathcal{L}  = L +iP ,
\end{equation}
with $L$ and $P$ also being real functions. 
Insertion of \eqref{eq:phi_inf} into \eqref{eq:def_J} yields 
\begin{equation}
    \mathcal{J}  = \frac{2(\varphi_{[1]}\psi_{[2]} - \varphi_{[2]}\psi_{[1]})}{\lambda^4}
    +O(\lambda^{-5}) ,
\end{equation}
and subsequent
integration of \eqref{hyp_ev} gives the asymptotic solution

\allowdisplaybreaks
\begin{widetext}
\begin{subequations}
    \begin{align}
        r(   w , \lambda) =& H \lambda + R_\infty  - \left(\frac{\kappa H }{4}\right)\frac{(\varphi_{[1]}^ 2+\psi_{[1]}^ 2)}{\lambda} + O(\lambda^{-2})\;,\label{eq:r_inf}\\
        Q(   w , \lambda) 
        =& 
        Q_\infty  
        + \frac{4\pi qH^2 (\varphi_{[1]}\psi_{[2]} - \varphi_{[2]}\psi_{[1]})}{\lambda} + O(\lambda^{-2}) \;,\\
        \alpha(   w , \lambda) 
        =& 
        \alpha_\infty  
         +\left(\frac{\epsilon Q_\infty}{H^2}\right)\frac{1}{\lambda} + O(\lambda^{-2}) \;,\\
        Z(   w , \lambda) 
        =&  
        Z_\infty
        + \frac{\kappa  Q^2_\infty}{8\pi H^2}\frac{1}{\lambda}
        + O(\lambda^{-2})\;,\label{eq:Zinfijn}
        \\
        L(   w , \lambda) 
                =& 
        L_\infty  
        -\frac{\epsilon \varphi_{[1]} }{H\lambda}
        +q\frac{
           \epsilon Q_\infty \psi_{[1]} +2\alpha_\infty H(R_\infty\psi_{[1]}+H \psi_{[2]})  }{H\lambda} +O(\lambda^{-2})\;,\\   
        P(   w , \lambda) 
                =& 
        P_\infty  
        -\frac{\epsilon \psi_{[1]}}{H\lambda}
        -q\frac{ 
          \epsilon Q_\infty \varphi_{[1]} +2\alpha_\infty H(R_\infty\varphi_{[1]}+H\varphi_{[2]})  }{H\lambda} 
        +O(\lambda^{-2})\;,\\        %
        V(   w , \lambda) 
        =&
        U_\infty  \lambda + V_\infty 
        +\frac{2(HZ_\infty -R_\infty)+\epsilon\kappa H^ 2(L_\infty \varphi_{[1]}+P_\infty\psi_{[1]})}{2H^ 3\lambda}
        +O(\lambda^ {-2})\;,
        \label{eq:V_inf} 
        \end{align}
\end{subequations}
\end{widetext}
where $H$, $R_\infty$, $Q_\infty$, $\alpha_\infty$, $Z_\infty$, $L_\infty$, $P_\infty$, $U_\infty$ and $V_\infty$ are functions of integration depending on $w$.
 Using \eqref{eq:qABEAB}, we relate the functions integration $U_\infty$ and $V_\infty$, 
\begin{eqnarray}
    U_\infty&=&
    -\frac{2\epsilon H^\prime}{H}\label{eq:Uinf}\;,\\
    V_\infty&=&
    \frac{1 
    -2\epsilon HR_\infty^\prime}{H^2}\;\;,
    \label{eq:Vinf}
\end{eqnarray}
where $^\prime$ denotes the total derivative with respect to the single independent argument of the respective function.
The evolution equations \eqref{ev_eqn_Phi} relate the constants $L_\infty$ and $P_\infty$  to the $w$ derivatives of $\varphi_{[1]}$ and $\psi_{[1]}$, respectively,
\begin{eqnarray}
\label{eq:LP1}
L_\infty&=&2(H \varphi_{[1]})^\prime\;,\\
\label{eq:LP2}
P_\infty&=&
2(H \psi_{[1]})^\prime\;.
\end{eqnarray}
The Bondi mass \eqref{eq:Bondi_mass} can be found from the limit of the Misner-Sharp mass \eqref{eq:misner_sharp_mass}
\begin{align}\label{eq:mbondir-zinfy}
m_B=\frac{1}{2}(R_\infty-HZ_\infty)\;.
\end{align}

From this result, an alternative calculation of the Bondi mass using the hypersurface variables is 
\begin{equation}\label{mB_limit}
m_B = -\frac{1}{2}\lim_{\lambda\rightarrow\infty}\left[\lambda^2\left(\frac{r}{\lambda}\right)_{,\lambda}+HZ\right]\;\;.
\end{equation}

The metric for large values of the affine parameter becomes
\begin{equation}\label{eq:asympt_metric}
\begin{split}
ds^2 =& -\left[-\frac{2H^ \prime \lambda}{H} + \frac{1-2\epsilon H R^\prime_\infty}{H}\right]dw^2 +2\epsilon dwd\lambda \\
&+ (H\lambda)^ 2q_{AB}dx^Adx^B 
+O(\lambda^{-1})\;.
\end{split}
\end{equation}
The asymptotic   transformation 
\begin{equation}
    x(w, \lambda) = H(w)\lambda + R_\infty(w) + O(\lambda^{-1}),
\end{equation}
with $w$ and the angles kept unaltered 
casts  
\eqref{eq:asympt_metric} into
\begin{equation}
\begin{split}
    ds^2 = &- \left(\frac{dw}{H}\right)^2
    +2\epsilon dx \left(\frac{dw}{H}\right) 
    \\
    &
    + x^2
    \left[1+  O(x^{-2})\right]^2
    q_{AB}dx^Adx^B  \;\;.
\end{split}
\end{equation}
Then if  we define $dw_B = dw/H$ or alternatively 
\begin{equation} \label{eq:BondiTime}
 \frac{d w_{B}}{d w} =\frac{1}{H}\;,
\end{equation}
the metric \eqref{eq:metric} approaches 
\begin{equation}\label{eq:asympt_inertial}
    ds^2 = -(dw_{B})^2
    +2\epsilon dw_{B}dx 
    + x^2 q_{AB}dx^Adx^B + O(x^{-1})\;\;,
\end{equation}
which is the metric of an inertial observer for large values  of the affine parameter. 
The coordinate frame in which $H=1$ is called the  Bondi frame and $w_B$ is referred to as the Bondi time,  corresponding to the inertial time of this observer.  
Consequently, in a regime where $H\neq0$, it measures the redshift between asymptotic inertial observers and observers at finite affine parameter distance. 
In particular, if $H\neq0$, we can always find a coordinate transformation of the type \eqref{eq:BondiTime} such that $H=1$. 
If, however,  $H=0$ for some time $w=w_E$, then $w_E$ corresponds to the time of formation of an event horizon, since the  surface forming rays of the  {hypersurface $w=$const} do not have an  end point in the asymptotic region. 
The redshift can also conveniently be calculated via
\begin{equation}
H = \lim_{\lambda\rightarrow\infty} r_{,\lambda}.
\end{equation}
In a Bondi frame, the metric fields behave as
\begin{equation}
\begin{split}
    r&= \lambda + R_\infty(w_B) + O(\lambda^{-1}),\\
    V&= 1 - 2\epsilon R_\infty^\prime(w_B)
    +
    O(\lambda^{-2}),\\
\end{split}
\end{equation}
and  the scalar field $\Phi$ admits the expansion,
\begin{equation}
\Phi = \frac{\mathcal{C}(w_B)}{r}+\frac{\mathcal{C}_{NP}}{r^2} +O(r^{-3}),
\end{equation}
where $\mathcal{C}(w_B)=\mathcal{C}_\varphi(w_B) + i\mathcal{C}_\psi(w_B)$ is the scalar monopole and $\mathcal{C}_{NP} $ is the conserved Newman-Penrose constant. 
These two quantities can alternatively be  determined via the limits
\begin{eqnarray}
\mathcal{C} &=& \lim_{r\rightarrow\infty } r\Phi  
=\lim_{\lambda \rightarrow\infty } H\lambda \Phi \nonumber\\
&=&  H (\varphi_{[1]} + i\psi_{[1]}), \label{eq:monopole}\\
\mathcal{C}_{NP}&=&-\!\lim_{r\rightarrow\infty } r^2(r\Phi)_{,r}  
=-\!\lim_{\lambda\rightarrow\infty } \frac{r^2(r\Phi)_{,\lambda}}{r_{,\lambda}} 
\nonumber\\
&=& R_\infty \mathcal{C}+H^ 2(\varphi_{[2]}+i\psi_{[2]}).
\end{eqnarray}
The derivative
\begin{equation}\label{eq:def_news}
\mathcal{N}(w_B) := \frac{d\mathcal{C}}{dw_B},\;\;
\end{equation}
is the  {\it  news function} of the scalar field, whose naming will become justified below.
The relation between the Bondi time $w_B$,  and the asymptotic time $w$ is given by asymptotic transformation \eqref{eq:BondiTime};
therefore,
\begin{equation}
\mathcal{N}(w_B) := \frac{d\mathcal{C}}{dw}\frac{dw}{dw_B}
=H\frac{d \mathcal{C}}{dw}.
\end{equation}
The asymptotic limit of the Weyl scalar $\Psi_2$ gives
\begin{equation}
\begin{split}   
\lim_{\lambda\rightarrow \infty}\lambda^3\Psi_2 =& 
\frac{2HZ_\infty - R_\infty}{2H^3}  \\
& +\frac{\kappa \epsilon}{6H^2}  \left[H\varphi_{[1]}\frac{d}{dw}H\varphi_{[1]}+H\psi_{[1]}\frac{d}{dw}H\psi_{[1]}\right],  
\end{split}
\end{equation}
in a Bondi frame ($H=1$, $r=\lambda$), and we can deduce  that
\begin{equation}
\begin{split}
    \lim_{r\rightarrow \infty}\Psi_2 =& -m_B(w_B)  \\
    &+\frac{1}{12}\kappa \epsilon\left[\mathcal{C}(w_B)\bar{\mathcal{N}}(w_B)+\bar{\mathcal{C}}(w_B)\mathcal{N}(w_B)\right]\;,
\end{split}
\end{equation}
which gives an alternative means to calculate the Bondi mass in a Bondi frame
\begin{equation}\label{eq:bondi_mass_psi2}
m_B = -\lim_{r\rightarrow \infty}\left\{ r^3\Psi_2 
+ \frac{1}{12}\kappa \epsilon r^2\left[\Phi\partial_w \bar{\Phi}+\bar{\Phi}\partial_w \Phi\right]\right\}.
\end{equation}
The expression \eqref{eq:bondi_mass_psi2} corresponds to an expression presented in \cite{Bicak:2010tt}.  
Evaluation of the supplementary equations \eqref{eq:E0} and \eqref{eq:E00} in a Bondi frame ($H=1$) gives us the asymptotic balance laws 
\begin{widetext}
\begin{subequations}\label{eq:BM_Q_conservation_component}
\begin{eqnarray}
   \frac{d m_B}{dw_B}  
   &=& 
   \frac{\epsilon\kappa}{2} \left\{ \left(\frac{d\varphi_{[1]}}{dw_B}\right)^2
   +\left(\frac{d\psi_{[1]}}{dw_B}\right)^2  
   +2 q\alpha_\infty \left[\phi_{[1]}\left(\frac{d\psi_{[1]}}{dw_B}\right) - \psi_{[1]}\left(\frac{d\phi_{[1]}}{dw_B}\right)\right]
   +q^ 2\alpha_\infty^ 2\left(\phi_{[1]}^ 2+ \psi_{[1]}^ 2\right)\right\},
    \\
    \frac{d Q_\infty}{dw_B}  
    &=&
    -4\pi q^ 2 \alpha_\infty^ 2\left(\phi_{[1]}^ 2+ \psi_{[1]}^ 2\right)
    +4\pi q\left[\phi_{[1]} \left(\frac{d\psi_{[1]}}{dw_B}\right)- \psi_{[1]}\left(\frac{d\phi_{[1]}}{dw_B}\right)\right],
\end{eqnarray}
\end{subequations}
\end{widetext}
which we can write also in terms of the complex news function and complex scalar monopole
\begin{subequations}\label{eq:BM_Q_conservation}
\begin{eqnarray} 
   \frac{d m_B}{dw_B}  
   &=& 
   \frac{\epsilon\kappa}{2}\left|  \mathcal{N} + 
    i  q\alpha_\infty  \mathcal{C}\right|^2,
    \\
    \frac{d Q_\infty}{dw_B}  
    &=&
    -4\pi q^ 2 \alpha_\infty^ 2|\mathcal{C}|^ 2
    -2i\pi q \left(\mathcal{N}\bar{\mathcal{C}}-\bar{\mathcal{N}}\mathcal{C}  \right).
\end{eqnarray}
\end{subequations}
As mentioned in Sec.~\ref{sec:fieldEQ}, there is a  gauge transformation to set\footnote{Since $\alpha_\infty=\alpha_\infty(w_B)$ this gauge transformation needs to be done  on each null hypersurface $w_B=$const for every value of $w_B$. } $\alpha_\infty=0$. 
Using $\alpha_\infty=0$ in \eqref{eq:BM_Q_conservation}  {casts the  Bondi} mass conservation law and the asymptotic  charge conservation law  into the simple expressions
\begin{equation}\label{eq:conservationLaws_BF}
    \frac{d m_B}{dw_B} = 
   \frac{\epsilon\kappa}{2}   |\mathcal{N}|^ 2\;\;,\;\;
   \frac{d Q_\infty}{dw_B}  
    =-2i\pi q \left(\mathcal{N}\bar{\mathcal{C}}-\bar{\mathcal{N}}\mathcal{C}  \right)\;\;.
\end{equation}
From the definition of the news function $\mathcal{N}$ in \eqref{eq:def_news} as the  derivative of the scalar monopole $\mathcal{C}$ and \eqref{eq:conservationLaws_BF}, we deduce that  the  asymptotic initial value problem for the charged version ($q\neq0$ ) requires specification of  three scalar values and one arbitrary function depending on $w_B$ while the uncharged version $(q=0)$ only requires  specification of two scalars and one function depending on $w_B$. 
In both cases, the free function is the  news function $\mathcal{N}$. 
Its name arises  from the same motivation as in the gravitational case ( see  \cite[Sec.~3]{Bondi:1962px}), and  the function $\mathcal{N}$ determines the complete (temporal) behavior of all the other physically measurable fields, $\mathcal{C}(w_B)$, $m_B(w_B)$ and $Q_\infty(w_B)$.

Both of the two versions of these asymptotic initial value problems can be treated together. 
 {Given a scalar news function $\mathcal{N}$ for $w_B\neq 0$, the three scalars  are specified at the initial time $w_B=0$. }
They are  $\mathcal{C}(0)$, $m_B(0)$ and $Q_\infty(0)$, which are the initial scalar monopole, the initial Bondi mass and the initial value of the global charge, respectively. 
The behavior of these fields for $w_B\neq 0$ is then determined by the following hierarchy of ordinary differential equations of  the asymptotic initial value problem
\begin{subequations}\label{eq:asymp_iv}
    \begin{eqnarray}
        \frac{d \mathcal{C}}{dw_B}&=&\mathcal{N}(w_B)\;\;,\\\frac{d m_B}{dw_B}&=&\frac{\epsilon \kappa}{2}|\mathcal{N}(w_B)|^2\;\;,\\
        \frac{d Q_\infty}{dw_B}&=&-2i\pi q \left[\mathcal{N}(w_B)\bar{\mathcal{C}}(w_B)-\bar{\mathcal{N}}(w_B)\mathcal{C}(w_B)  \right]\;\;.\nonumber\\
    \end{eqnarray}
\end{subequations}
In the uncharged version, where the gauge charge $q$ vanishes, the last equation of \eqref{eq:asymp_iv} implies $Q_\infty(w_B) = 0$ as the initial value of the global charge  $Q_\infty(0)$ is obviously zero. Thus only two initial values and one free function are required for the asymptotic initial value problem in this case.

\subsection{Local initial-boundary formulation at a world line}\label{sec:cibvp_worldline}
In spherical symmetry there exists a designated world line, which is the  geodesic tracing the centers of spherical symmetry  \cite{1960rgt..book.....S}. We denote this geodesic with $c(w)$ and choose $w$ as an  affine parameter for its parameterization. 

The tangent vector $c^a\partial_a$ is a unit timelike vector and $w$ measures the proper time along $c(w)$. 
This proper time is assumed to be the proper time of a freely falling Fermi observer along $c(w)$ \cite{Manasse:1963zz,Poisson:2009pwt}.
Following \cite{Madler:2012sg}, in a convex normal neighborhood of $c(w)$, we extend the definition of $w$ to be constant along null cones emanating from $c(w)$.
The surface forming rays of the $w=$const null cones are labeled by the angular coordinates $x^A$ within each null cone.
The null condition on $w=$const implies $g^{ww}=0$, since $g^{ab}\nabla_a w\nabla_b w=0$.
The covector $\ell_a=\epsilon\nabla _a w$ defines by metric duality the tangent vectors $\ell^a = \epsilon g^{ab}\nabla_b w = \epsilon g^{aw}$ of the surface forming null rays. 
The null rays are labeled by the angular coordinates, $\ell^a\nabla_a x^A=0$ and therefore $g^{wA}=0$.

\begin{figure*}[htbp]
  \centering
  \includegraphics[width=0.4\linewidth]{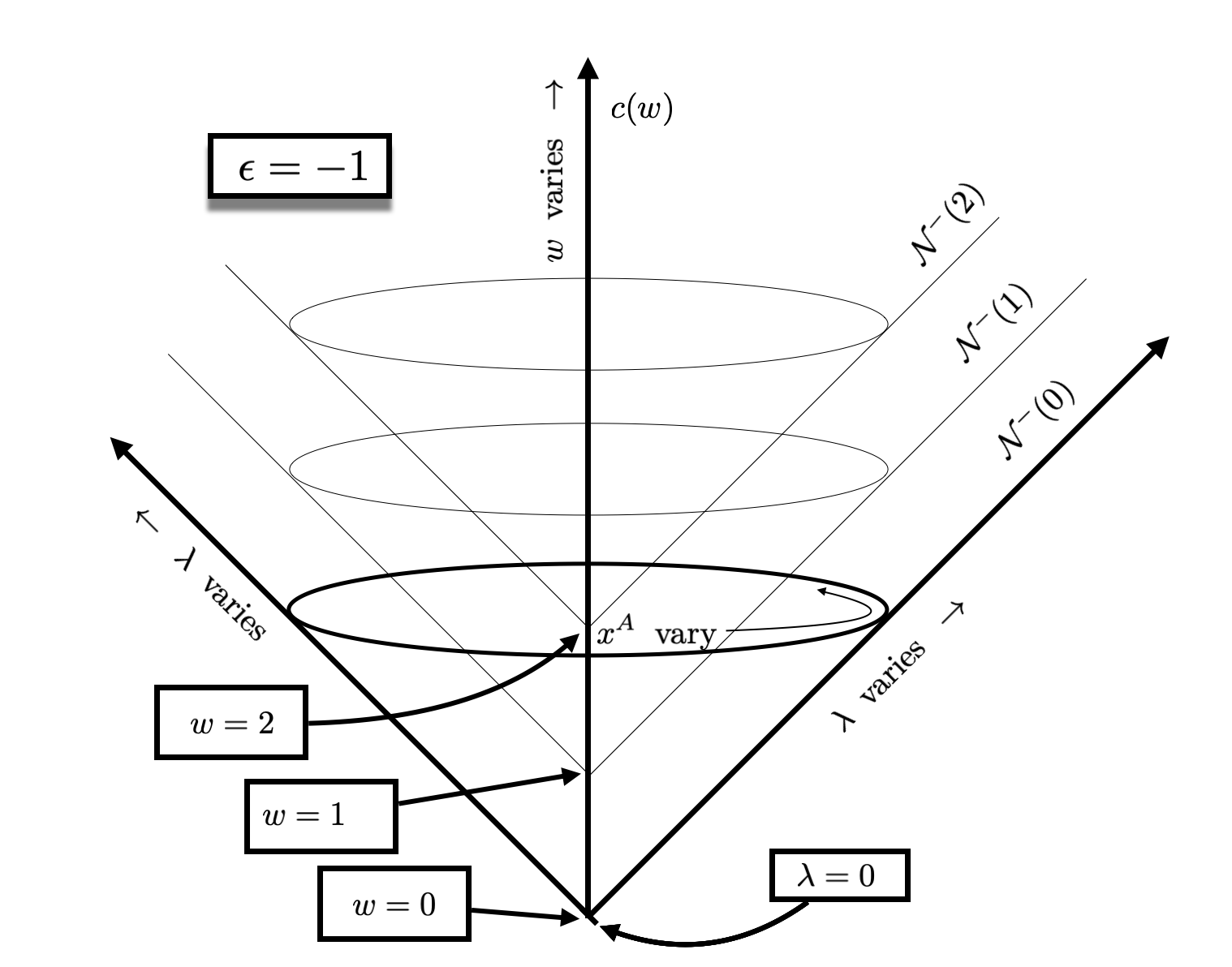}
  \includegraphics[width=0.4\linewidth]{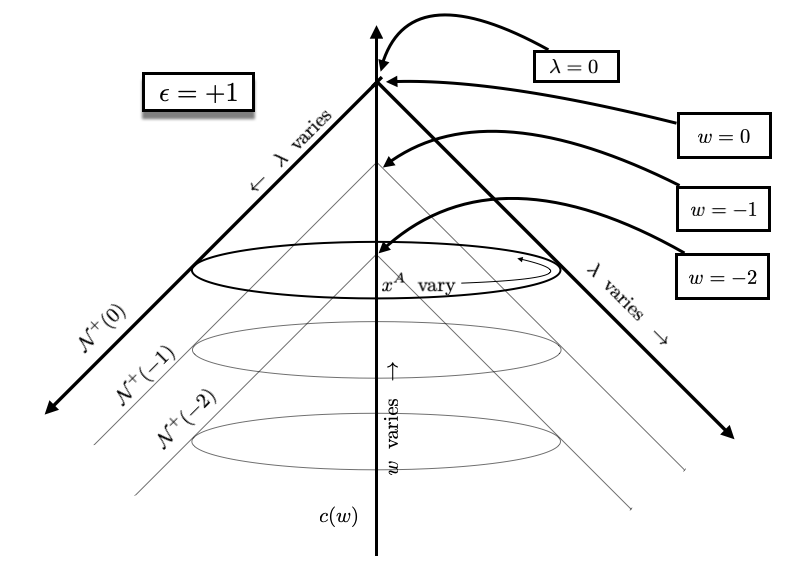}
  \caption{ {A family of null cones $\mathscr{N}^\epsilon(w)$ at the central geodesic  $c(w)$ is shown for $\epsilon=-1$ (left) and $\epsilon=+1$ (right). The left graphic shows the local geometry for the characteristic initial value problem where initial data are provided in the null cone $\mathscr{N}^-(0)$ and the right graphic illustrates it for the characteristic past value problem with  the future  data on  $\mathscr{N}^+(0)$. The origin of the affine parameter $\lambda=0$ for each null cone $\mathscr{N}^-(w)$ is at the vertices of the  cones  traced by  $c(w)$. Distinctive  vertices are  labeled with different values of $w$. }}
  \label{fig:worldline_null}
\end{figure*}

By  these coordinate conditions, only the coordinate $x^1=\lambda$ varies along the rays of a given null cone $w=$const. 
Since the cones emanate from $c(w)$, we take the origin $\lambda=0$ to be on $c(w)$.
We normalize the parameter $\lambda$ along the rays such that it is an affine parameter with the normalization determined at $c(w)$.
In particular, $c^a\nabla_a\lambda = \ell^a\nabla_a \lambda = \epsilon g^{w\lambda} = 1$ with $|\epsilon|=1$ at $c(w)$ so that $\epsilon=-1$ designates that the respective null cone $w=$const are outgoing null cones with null rays pointing into the future of $c(w)$ and $\epsilon=1$ are ingoing null cones pointing into the past of $c(w)$ (see Fig.~\ref{fig:worldline_null}).

By these conditions, the metric near $c(w)$ has the form \cite{Madler:2012sg}\footnote{Reference \cite{Madler:2012sg} uses a reversed nomenclature for the binary parameter $\epsilon$.} 
\begin{equation}
\begin{split}
ds^2 =& -\left[1+O(\lambda^2)\right] (dw)^2 +2\epsilon dwd\lambda\\
&+ \lambda^2 q_{AB}dx^Adx^B + O(\lambda^3).
\end{split}
\end{equation}

This implies the regularity and boundary conditions for $\lambda=0$
\begin{equation}\label{eq:reg_WL}
\begin{split}
&r(w, 0) = 0\;\;,\;\;
r_{,\lambda}(w, 0) = 1\;\;,\;\;\\
&V(w, 0) = 1\;\;,\;\;
V_{,\lambda}(w, 0) = 0\;\;,\;\;
\end{split}
\end{equation}
which also imply 
\begin{equation}\label{eq:bc_vertex_ZL}
Z(w, 0) =\mathcal{L}(w,0)=0 \;\;,\;\;
\end{equation}
The supplementary equation \eqref{eq:E00} holds at $c(w)$ due to the regularity conditions. Equations \eqref{eq:reg_WL} and \eqref{eq:E0} imply
\begin{equation}
Q_{,w}  = O(\lambda^2)\;\;\Rightarrow \;\;Q (w, 0) =Q_{,\lambda} (w, 0) = 0.
\end{equation}
At the vertex, we assume that the scalar field $\Phi$ has the regular Taylor expansion
\begin{equation}\label{eq:data_vertex}
\Phi  = \Phi_{(0)}  + \Phi_{(1)} \lambda + \frac{1}{2}\Phi_{(2)}\lambda^2 + O(\lambda^3)\;\;,
\end{equation}
where the subscript $(k)$ in parentheses indicate that the symbol carrying  the subindex is the $\lambda^{k}$ coefficient function  depending on $w$ of the Taylor series of the respective field evaluated at $\lambda=0$.

The scalar field expansions \eqref{eq:data_vertex} and \eqref{eq:def_J} give 
\begin{equation}
    \begin{split}
        \mathcal{J}  = & \mathcal{J}_{(0)} 
        + \mathcal{J}_{(1)}\lambda + O(\lambda^2)\\
        =&i\left(\overline {\Phi}_{(1)} \Phi_{(0)}
        -\overline {\Phi}_{(0)} \Phi_{(1)} \right)
        +i\left(\overline {\Phi}_{(2)} \Phi_{(0)}
        -\overline {\Phi}_{(0)} \Phi_{(2)}\right)\lambda \\&+O(\lambda^2)\;.
    \end{split}
\end{equation}

Insertion of these data into \eqref{hyp_ev} and integration give
\begin{widetext}
\begin{subequations}\label{eq:vertex_fields}
    \begin{eqnarray}
        r(w, \lambda) 
        &=& 
        \lambda -\frac{\kappa \lambda^3}{12}|\Phi_{(1)}|^2 -\frac{\kappa \lambda^4}{24}\Bigl(\Phi_{(2)} \overline{\Phi}_{(1)}+\Phi_{(1)} \overline{\Phi}_{(2)}\Bigr) + O(\lambda^5)
        \;,
        \label{eq:sol_r_vertex}\\
        Q(w, \lambda) 
        &=& 
        - \pi q  \left( \frac{2}{3} \mathcal{J}_{(0)} 
        + \frac{1}{2}   \mathcal{J} _{(1)}\lambda\right)\lambda^3 + O(\lambda^5)\;,\label{eq:sol_Q_vertex}\\
        \alpha(w, \lambda) 
        &=& \alpha_0 
    +\frac{\pi \epsilon q}{3}  \left(\mathcal{J}_{(0)}
    +\frac{1}{2}\mathcal{J}_{(1)}\lambda\right) \lambda^2
    + O(\lambda^4)\;,\label{eq:sol_alpha_vertex}\\
    Z(w,\lambda)&=&
    -\frac{\pi \kappa  q^2 \mathcal{J}_{(0)}}{90}  \Big( \mathcal{J}_{(0)} +\frac{5}{4}\mathcal{J}_{(1)}\lambda \Big) \lambda^5 + O(\lambda^7)\;,\\
    \mathcal{L}(w,\lambda)&=&
    -\left(\epsilon \Phi_{(1)}+2iq\alpha_0\Phi_{(0)}\right)\lambda 
    -\left(\frac{\epsilon \Phi_{(2)}}{2}+2iq\alpha_0\Phi_{(1)}\right)\lambda^2 
    + O(\lambda^3)\;,\\
    V(w,\lambda)&=&
    1+\frac{\kappa}{3}|\Phi_{(1)}|^2\lambda^2 
    +O(\lambda^3)\;,\label{eq:sol_V_vertex}\\
    \Phi_{,w}&=&
     -\epsilon\Phi_{(1)}-i q\alpha_0  \Phi_{(0)}
    -\left[\frac{3}{4}\epsilon\Phi_{(2)} +iq\alpha_0 \Phi_{(1)}
    \right]\lambda+ O(\lambda^2)\;.\label{eq:dphidw_vertex}
    \end{eqnarray}
\end{subequations}
\end{widetext}
The gauge freedom of the electromagnetic field allows us in principle to set $\alpha_0$ everywhere along $c(w)$.
Moreover, from \eqref{eq:dphidw_vertex} it is not difficult to deduce that
\begin{align}\label{eq:dphidw_vertex}
\Phi_{(0),w} =& -\epsilon\Phi_{(1)}-i q\alpha_0  \Phi_{(0)}\;,\\ \quad
\Phi_{(1),w} =& -\frac{3}{4}\epsilon\Phi_{(2)} -iq\alpha_0 \Phi_{(1)}\;.
\end{align}
 
 We can see that the time evolution of the coefficient $\Phi_{(i)}$ along  $c(w)$ is driven by a function depending on the expansion coefficients containing all expansion coefficients up to $\Phi_{(i+1)}$. 
  {This behavior is a result} of the choice of $w$ being constant along a null hypersurface. 
The hierarchical structure of \eqref{eq:dphidw_vertex} is the scalar field equivalent to the similar structure in the gravitational case \cite{Madler:2012sg,Madler:2012ovw} for gravitational waves.
Since $\Phi_{(1)} = \lim_{\lambda\rightarrow 0} \Phi_{,\lambda}$ and $\Phi_{(2)} = \lim_{\lambda\rightarrow 0} \Phi_{,\lambda\lambda}$, we learn from \eqref{eq:vertex_fields} that the evolution of the initial data $\Phi$ along $c(w)$ and in its neighborhood is completely determined by the data on a cone $w=$const, provided the data and its derivatives are finite in the limit $\lambda\rightarrow 0$. 
Indeed, it is not difficult to derive similar equations like in \eqref{eq:dphidw_vertex} for $\Phi_{(k)}$ with $k\ge 2$, while using an appropriate higher order  expansion than \eqref{eq:data_vertex} and integrating \eqref{hyp_ev}. 

From such an exercise, we can deduce the following {\it local} CIBVP at the vertices of null cones emanating from a geodesic world line: Suppose, the complex scalar field $\Phi(w, \lambda)$ on a null cone $w=0$ has a regular Taylor series expansion, 
\begin{equation}
    \Phi(0,\lambda) = \sum^\infty_{k=0}\frac{\lambda^k}{k!}\frac{\partial^k\Phi}{\partial \lambda^k}\Bigg|_{\lambda=0}\;,
\end{equation}
evaluated at the world line, the boundary conditions of the fields are \eqref{eq:reg_WL} and  \eqref{eq:bc_vertex_ZL}, $\alpha_0$ is an arbitrary function of $w$ for $w\neq 0$, and the gauge charge $q$ is provided; then, the metric, the electromagnetic potential $\alpha$ and the charge function $Q$ can be completely calculated on the cone $w=0$ using the hypersurface equations \eqref{eq:hyp_r}-\eqref{eq:hyp_V}. Their local solution at $\lambda=0$ is of the form \eqref{eq:sol_r_vertex}, \eqref{eq:sol_Q_vertex}, \eqref{eq:sol_alpha_vertex}, or \eqref{eq:sol_V_vertex}.
The evolution equations \eqref{ev_eqn_Phi} determine the time derivative of the Taylor coefficients of the initial data, as seen in, e.g., \eqref{eq:dphidw_vertex}.

We remark, that this regularity assumption at $w=0$ does not imply  that the data necessarily remain regular for $w\neq 0$.
If $r\rightarrow 0$ for $\lambda\rightarrow\lambda_C>0 $ a (massive) caustic forms at finite value for the affine parameter. This  caustic formation for some $w=w_0\neq 0$ is in general driven  by  the value of the norm squared of the derivative of the scalar field, $|\Phi_{,\lambda}|^2$. 
An example for  {caustic initial data} was recently given in \cite{Madler:2024kks}, where the data were employed in an evolution through an event horizon of a supercritical scalar field. 
The  formulated initial value problem along a world line for a spherically symmetric charged scalar field is an analog to the gravitational CIBVP \cite{Friedrich:1986rb}.

\subsection{Initial-boundary value problem at a  null hypersurface}\label{sec:cibvp_dn}
\begin{figure*}[htbp]
  \centering
  \includegraphics[width=0.4\linewidth]{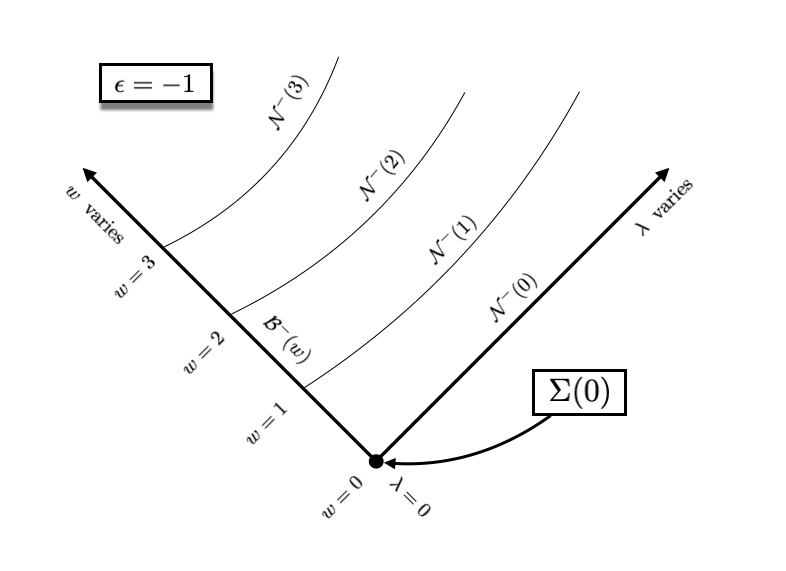}
  \includegraphics[width=0.4\linewidth]{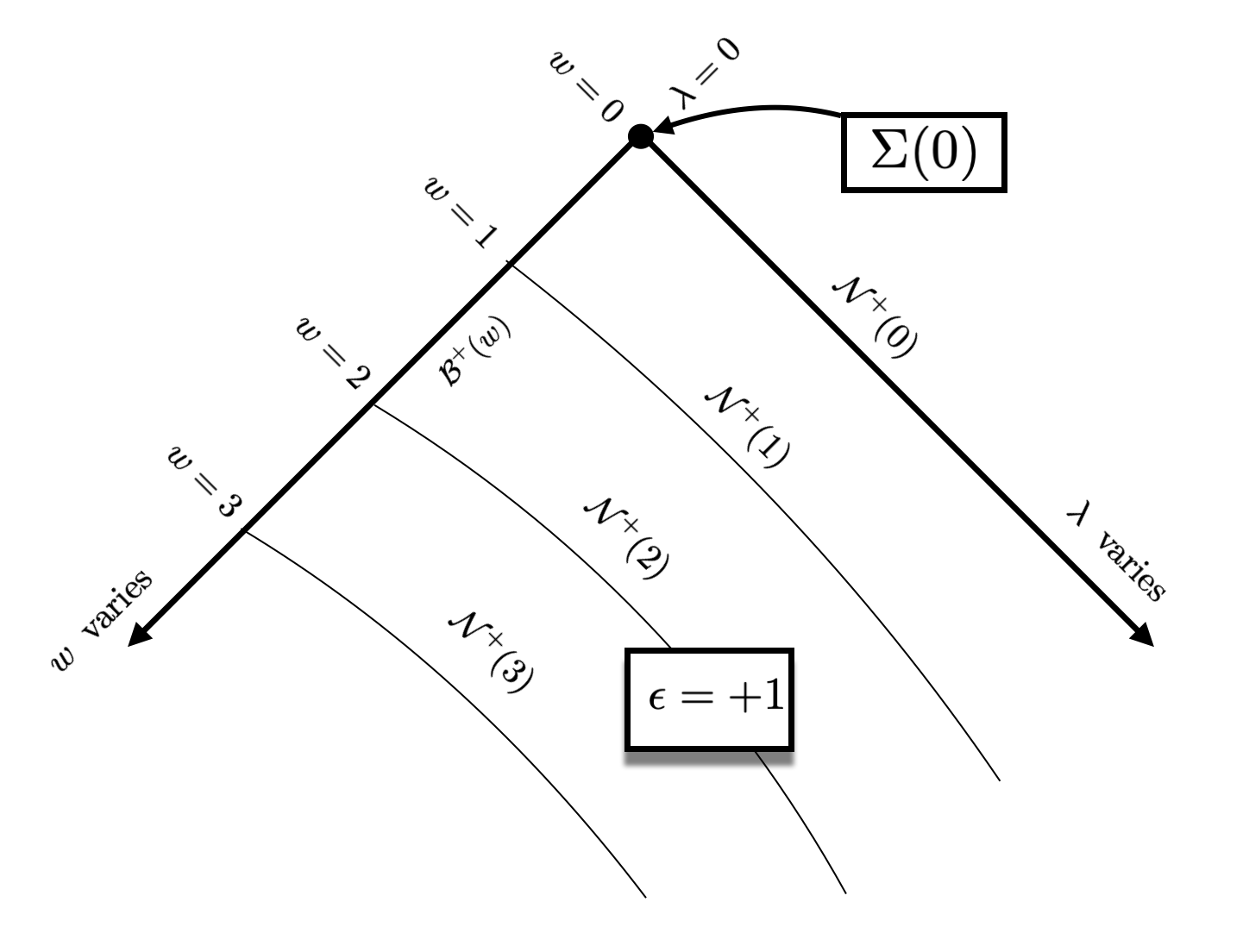}
  \caption{A family of null hypersurfaces $\mathscr{N}^\epsilon(w)$ with corresponding  boundary null hypersurface  $\mathcal{B}^\epsilon(w)$ is shown for $\epsilon=-1$ (left) and $\epsilon=+1$ (right). The former is a characteristic initial value problem with  the initial data surface $\mathscr{N}^-(0)$ and the latter gives rise to a characteristic past value problem with  the past  data on  $\mathscr{N}^+(0)$. The common intersection between $\mathscr{N}^\epsilon(0)$ and $\mathcal{B}^\epsilon(w)$ is $\Sigma(0)$.}
  \label{fig:bh}
\end{figure*}

\noindent 
Consider a family of  null hypersurfaces $w=$const designated by $\mathscr{N}^ \epsilon(w)$, hereafter. 
The generators $\ell^a\partial_a$ of these null hypersurfaces are affinely parametrized with the affine parameter $\lambda$.
We choose the origin, $\lambda=0$, of the affine parameter to be at a null hypersurface $\mathcal{B}^ \epsilon$. The surface $\mathcal{B}^ \epsilon$ is a null boundary for every null hypersurface $w=$const of $\mathscr{N}^ \epsilon$ and is the boundary surface for the double-null CIBVPs to be presented, 
see Fig.\eqref{fig:bh}.

Near $\mathcal{B}^ \epsilon$ on $\mathscr{N}^ \epsilon$, the metric must behave like
\begin{equation}
ds^2 =  O(\lambda) (dw)^2 +2\epsilon dwd\lambda + r^2(w, 0) q_{AB}dx^Adx^B + O(\lambda^3).
\end{equation}
This implies $V(w, 0) = 0$
for all values of $w$,  where $\lambda=0$. Consequently, $V = O(\lambda)$ near $\mathcal{B}^ \epsilon$ on $\mathscr{N}^\epsilon$. 
For the   conformal factor $r$ there are two possibilities regarding its evaluation on $\mathcal{B}^ \epsilon$: (i) $r(w, 0)=0$, and (ii) $r(w, 0)\neq0$.

Physically these cases mean that the null hypersurface $\mathcal{B}^\epsilon$ has either a trivial [case (i)] or nontrivial Misner-Sharp mass [case (ii)], which becomes clear considering \eqref{eq:misner_sharp_mass} in the limit $\lambda\rightarrow\infty$
\begin{equation}\label{eq:MSM_lambda0}
    \lim_{\lambda\rightarrow\infty} M
    =\frac{r(w,0)}{2}\left[1 - 2\epsilon r_{,w}(w,0)r_{,\lambda}(w,0) \right]\;\;.
\end{equation}

The two options lead to different implications for the formulation of an initial boundary value problem.
These options are considered separately, hereafter.
\subsubsection{Initial boundary formulation at a null hypersurface with $r(w, 0)=0$}\label{sec:dn_req0}
\noindent
If $V(w,0)=r(w,0)=0$, the line element near the origin of the affine parameter $\lambda$  can be written as
\begin{equation}\label{eq:ds2_at_Bdn_1}
\begin{split}
ds^2 =&  -\left[V_{(1)}(w)\lambda + O(\lambda^2) \right](dw)^2 +2\epsilon dwd\lambda\\
&+ [r_{(1)}(w)\lambda]^2\left[1+O(\lambda)\right] q_{AB}dx^Adx^B\;,
\end{split}
\end{equation}
where it is required that $r_1\neq0$. Locally there exists a coordinate transformation, $\hat \lambda = r_1\lambda$ and $\hat w = \int (dw/r_1)$, so that the line element takes the form
\begin{equation}
\begin{split}
    ds^2 =&  -\left\{\left[r_{(1)}V_{(1)} + 2\epsilon r_{(1),\hat w} \right] \hat \lambda +O(\hat\lambda^2)\right\}(d\hat w)^2  \\
    &+2\epsilon d\hat wd\hat \lambda
    + \hat \lambda ^2\left[1+O(\hat \lambda)\right] q_{AB}dx^Adx^B.
\end{split}
\end{equation}

Consequently, we can set wlog $r_{(1)}=1$, which is assumed hereafter in this subsection.
With the assumption of a regular expansion of the scalar field as in \eqref{eq:data_vertex}, we find that $Z(w, 0)=\mathcal{L}(w,0)=0$. The residual gauge freedom of the electromagnetic field $\alpha$ allows us to set $\alpha(w, 0)=0$.  In addition, the supplementary equation \eqref{eq:E0} implies $Q(w, 0)$=const for $\lambda=0$ and we set wlog $Q(w,0)=0$. 
The  supplementary equation \eqref{eq:E00} gives in the limit $\lambda\rightarrow 0$
\begin{equation}\label{eq:V1_evolution}
0= V_{(1),w} +\epsilon \left[-V_{(1)}^2 + \kappa |\Phi_{(0),w}|^2\right].
\end{equation}

This equation can be solved along the null boundary $\mathcal{B}^\epsilon$ for  values $w\neq 0$, provided an initial value $V_{(1)}(0)$ is given at $w=0$ and provided the function $\Phi_{(0),w}$ is known for all values $w\ge0$. 
The function  $V_{(1)}$ measures the inaffinity of the null geodesics $n^a\partial_a$ generating the null hypersurface $\mathcal{B}^\epsilon$, which follows from \eqref{eq:n_inaffine}.
For a nonvanishing and $w$-dependent scalar field on $\mathcal{B}^\epsilon$, \eqref{eq:V1_evolution} shows that the evolution of the inaffinity along $\mathcal{B}^\epsilon$ is driven by the free function $\Phi_{(0),w}$.

Thus, the function $\Phi_{(0),w}$ can be seen as some kind of analog of the news function at $\lambda=0$ instead of in the asymptotic limit $\lambda\rightarrow\infty$.

A CIBVP can now be constructed in the following way. Let $w=0$ be a null hypersurface $\mathscr{N}^ \epsilon(0)$ whose generators are tangent to affinely parametrized null geodesics with tangent $\ell^a\partial_a$ and affine parameter $\lambda$.  
At the origin of the affine parameter, $\lambda=0$, the null hypersurface $\mathscr{N}^ \epsilon(0)$ intersects a null hypersurface $\mathcal{B}^\epsilon$ which has the generators $n^a\partial_a$ that obey the normalization $n^a\ell_a = -1$. 
The intersection of the two null hypersurfaces $\mathscr{N}^\epsilon(0)$ and $\mathcal{B}^\epsilon$ is a two-dimensional spacelike space with spherical topology and unit Gaussian curvature, called $\Sigma$. 
The intrinsic metric of $\Sigma$ is the one of the unit sphere, $q_{AB}$. 
For this double-null CIBVP, the following specification of data determines  a solution of the Einstein-scalar field equations in spherical symmetry:
\begin{itemize}
    \item specify the type of null hypersurface $w=0$, by choosing either $\epsilon=1$ (ingoing null hypersurface) or $\epsilon=-1$ (outgoing null hypersurface);
    \item specify the gauge charge $q$;
     \item on $\mathscr{N}^ \epsilon(0)$ specify a free complex function $\mathcal{F}_{\Phi}(\lambda)$ that can be expressed by  a regular Taylor series at $\lambda=0$;
    \item on $\mathcal{B}^\epsilon$ specify a free complex function $\mathcal{N}_{\mathcal{B}}(w)$; 
    \item on $\Sigma$ specify the following fields: 
    $\Phi(0,0)$ and $V_{(1)}(0)$;
    \item on $\mathcal{B}^\epsilon$, i.e. for $\lambda=0$, the following fields are trivial: $r$, $V$, $Q$, $\alpha$, $Z$  and $\mathcal{L}$;
    \item on $\mathcal{B}^\epsilon$, the field $r_{,\lambda}$ is nontrivial, i. e. $\lim _{\lambda\rightarrow0} r_{,\lambda} =1$ for all values $w\ge0$.
\end{itemize}
With these specifications, the field equations are solved in the following way:
\begin{enumerate}
    \item With  $\mathcal{N}_{\mathcal{B}}(w)$ and $\Phi(0,0)$ determine the scalar field everywhere on $\mathcal{B}^ \epsilon$ via the equation
    \begin{equation}
        \Phi_{,w}(w, 0) = \mathcal{N}_{\mathcal{B}}(w).
    \end{equation}
    \item With  $\mathcal{N}_{\mathcal{B}}(w)$ and $V_{(1)}(0)$ determine the inaffinity everywhere on $\mathcal{B}^ \epsilon$ via the equation [which follows from \eqref{eq:V1_evolution}]
    \begin{equation}
        V_{(1),w} =-\epsilon \left[V_{(1)}^2 + \kappa |\mathcal{N}_{\mathcal{B}}|^2\right].
    \end{equation}
    \item With  $\mathcal{F}_{\Phi}(\lambda)$ and $\Phi(0,0)$ determine the scalar field everywhere on $\mathscr{N}^ \epsilon(0)$ via the equation
    \begin{equation}
        \Phi_{,\lambda}(0, \lambda) = \mathcal{F}_{\Phi}(\lambda).
    \end{equation}
    \item As all boundary data for $r$, $Q$, $\alpha$, $Z$, $\mathcal{L}$ and $V$ are known on $\mathcal{B}^\epsilon$, the hypersurface equations \eqref{eq:hyp_r} - \eqref{eq:hyp_V} can be hierarchically integrated for  all the fields relevant to determine the $w$-derivative of $\Phi$ in Eq. \eqref{ev_eqn_Phi}
    \item Integration of the $w$-derivative of $\Phi$ by  Eq. \eqref{ev_eqn_Phi} (using, for example, a finite difference scheme)  determines $\Phi(\Delta w, \lambda)$ at some time $w=\Delta w\neq 0$ on a null hypersurface $\mathscr{N}^\epsilon(\Delta w)$
    \item the new values $\Phi(\Delta w, \lambda)$ serve as initial data on $\mathscr{N}^\epsilon(\Delta w)$ and are used to determine the metric and electromagnetic fields following step 4.
\end{enumerate}
This algorithm can be suitably implemented numerically. 
\subsubsection{Initial boundary formulation at a null hypersurface with $r(w, 0)\neq 0$}
If the areal distance function is nontrivial on the boundary $\mathcal{B}^\epsilon$, i.e., $r(w, 0)\neq 0$, a different  boundary value formulation will come about. Near the boundary the metric has the behavior
\begin{equation}\label{eq:ds2_at_Bdn_2}
\begin{split}  
ds^2 =& -\left[V_{(1)}(w)\lambda+O(\lambda^2)\right] dw^2 + 2\epsilon dw d\lambda 
\\
&+ \left[r_{(0)}(w) + r_{(1)}(w)\lambda  +O(\lambda^2)\right]^2 q_{AB}dx^Adx^B \;\;,
\end{split}
\end{equation}
with $r_{(0)}(w) = r(w, 0)\neq 0$.
Suppose that  there is a coordinate transform like  
\begin{equation}
    w = \hat w + G(\hat w)\hat \lambda +O(\hat \lambda^2)\;\;,\;\;
    \lambda = (\hat \lambda-r_{(0)})/r_{(1)} + O(\hat \lambda^2)\;\;.
\end{equation}
 that casts \eqref{eq:ds2_at_Bdn_2} into a form similar to the expansion of  \eqref{eq:ds2_at_Bdn_1}. If that were the case, the initial boundary value problem could be discussed as in the previous section. 
 It is, indeed, possible to find a coordinate transformation for \eqref{eq:ds2_at_Bdn_2}  such that the resulting metric has $g_{AB} =\hat \lambda^2 q_{AB} + O(\hat \lambda^3)$, but such a transformation gives $g_{\hat w\hat w}|_{\hat \lambda =0} = O(1)\neq 0$ in the new coordinates. This violates the null condition at the boundary $\hat \lambda=0$ and cannot be removed without requiring  $r_{(0),w}=0$.
Since the coefficient $r_{(0)}$  corresponds to a mass, which becomes clear considering \eqref{eq:MSM_lambda0}, we discard the  restriction $r_{(0),w}=0$ as it implies that null boundary $\mathcal{B}^\epsilon$ has a mass which remains constant with respect to a $w$ evolution.
Thus, having $r_{(0)}\neq 0$ on $\mathcal{B}^\epsilon$ indicates that $\mathcal{B}^\epsilon$ is a null hypersurface with a given mass. 
The choice of \eqref{eq:ds2_at_Bdn_2} with $r_{(0)}>0$ corresponds to choosing a spherically symmetric Gaussian null coordinate system \cite{Moncrief:1983xua} at $\mathcal{B}^\epsilon$  since $g_{AB}$ is positive definite on $\mathcal{B}^\epsilon$.  
In the next section it will be shown that $\mathcal{B}^\epsilon$ with $r_{(0)}\neq 0$ can be related to a horizon of a black hole.
Although there is no local  coordinate transformation to cast \eqref{eq:ds2_at_Bdn_2} into a form like \eqref{eq:ds2_at_Bdn_1}, it can still be simplified because $V_{(1)}$ can be set to zero everywhere on  $\mathcal{B}^\epsilon$. The null vector generating  $\mathcal{B}^\epsilon$ is $n^a\partial_a$ and the generating curves are nonaffine-null geodesics, which becomes clear when projecting \eqref{eq:n_inaffine} to $\mathcal{B}^\epsilon$ whose only component is \begin{equation}\label{eq:nw_inaffine}
    n^a\nabla _a n^w = -\frac{1}{2}V_{(1)}n^w.
\end{equation}
The coordinate transformation $w = w(\hat w)$ brings \eqref{eq:nw_inaffine} into the form
\begin{equation}\label{eq:nw_inaffine}
    n^a\nabla _a n^{\hat w} = -\left(\frac{d }{d\hat w}\ln\left|\frac{d w}{d\hat w}\right|+ \frac{1}{2}V_{(1)} \right)n^{\hat w}.
\end{equation}
If the coordinate transformation is chosen such that the term in the parentheses of \eqref{eq:nw_inaffine} vanishes,  the nonaffine-null geodesics ruling $\mathcal{B}^\epsilon$ become affine-null geodesics and   $g_{\hat w\hat w} = O(\lambda^2)$ near $\mathcal{B}^\epsilon$. 
This removal of the inaffinity is, in fact, a classical feature of nonaffine geodesics \cite{Schouten_book,1960rgt..book.....S,Poisson:2009pwt} and not restricted to null geodesics.
We make this choice in this section. In addition, we make the gauge choice $\alpha=0$ everywhere on $\mathcal{B}^\epsilon$.
Evaluation of the supplementary equations \eqref{eq:E00} and \eqref{eq:E0} on $\mathcal{B}^\epsilon$ yields
\begin{subequations}\label{eq:null_boundary_equations}
\begin{eqnarray}
0&=&
 \left(r_{,ww}   
+\frac{1}{2}\kappa r 
|\Phi_{,w}|^2 
  \right)\bigg|_{\mathcal{B}^ \epsilon}\;\;,\label{eq:dn_r_on_B}\\
  0&=&
\left[ Q_{,w}  
+2\pi q r^2  i(\overline {\Phi} \Phi_{,w} - \Phi   \overline{ \Phi}_{,w}) \right] \bigg|_{\mathcal{B}^ \epsilon}\;\;.\label{eq:dn_Q_on_B}
\end{eqnarray}
The two equations \eqref{eq:dn_r_on_B} and \eqref{eq:dn_Q_on_B} determine the value of $r$ and $Q$ everywhere on $\mathcal{B}^\epsilon$ provided the scalar field $\Phi$ and its $w$-derivative are known on  $\mathcal{B}^\epsilon$. A solution of \eqref{eq:hyp_r} also requires the specification of $r_{,\lambda}$ on $\mathcal{B}^\epsilon$. The evaluation of the main equation $q^{AB}E_{AB}$, from \eqref{eq:qABEAB}, while using $V_{,\lambda}|_{\lambda=0} = 0$ yields
\begin{eqnarray}\label{eq:dn_drdlambda_on_B}
0&=&
 \left[
  2\epsilon (r r_{,\lambda})_{,w} 
 -1
+\frac{\kappa }{8\pi }\frac{Q^2}{r^2}
\right]\bigg|_{\mathcal{B}^\epsilon}\;\;.
\end{eqnarray}
\end{subequations}
Equation \eqref{eq:dn_drdlambda_on_B} can be used to determine $r_{,\lambda}$ everywhere on $\mathcal{B}^\epsilon$ provided $r$  and $Q$ are known on $\mathcal{B}^\epsilon$ and $r\neq 0$ on $\mathcal{B}^\epsilon$. 
The initial values for solving the hypersurface equations \eqref{eq:hyp_Z} and \eqref{eq:hyp_L} for $Z$ and $\mathcal{L}$ follow from  their definitions on $\mathcal{B}^\epsilon$
\begin{equation}
Z\Big|_{\mathcal{B}^ \epsilon} = 2\epsilon (rr_{,w})\Big|_{\mathcal{B}^ \epsilon}\;\;,\;\;
\mathcal{L}\Big|_{\mathcal{B}^ \epsilon} =2(r\Phi_{,w})\Big|_{\mathcal{B}^ \epsilon}\;\;,
\end{equation}
which show that the initial values for $Z$ and $\mathcal{L}$ are known once $r$ and $\Phi_{,w}$ are known on $\mathcal{B}^\epsilon$. 
With the information collected above,  a double-null CIBVP can be formulated with a nontrivial conformal factor $r$ on the null boundary $\mathcal{B}^\epsilon$.

In a spherically symmetric spacetime, let $\mathscr{N}^\epsilon(w)$ be a family of  null hypersurface $w=$const with generators $\ell^a\partial_a$ that are affinely parametrized with an affine parameter $\lambda$. 
The affine parameter $\lambda$ is normalized with a  parameter $\epsilon$ via  $\ell^a\nabla_a \lambda = \epsilon$ with $\epsilon^2=1$.
The binary parameter $\epsilon$  indicates whether a null hypersurface $w=$const is an  outgoing null hypersurface ($\epsilon=-1$) or an ingoing null hypersurface ($\epsilon=1$). 
The null hypersurfaces $\mathscr{N}^\epsilon(w)$ intersect a null hypersurface $\mathcal{B}^\epsilon(w)$, where $\lambda=0$ on $\mathscr{N}^\epsilon(w)$. 
The generators of $\mathcal{B}^\epsilon(w)$ are  null vectors $n^a\partial_a$.  
These generators obey the cross normalization $\ell^an_a = -1$ and the integral curves (null geodesics) of $n^a$ are affinely parametrized with the parameter $w$ on $\mathcal{B}^\epsilon(w)$. The intersections of $\mathscr{N}^\epsilon(0)$ and $\mathcal{B}^\epsilon(w)$ are spatial spherical cross sections $\Sigma(w)$ of codimension-2 with an intrinsic two metric $g_{AB}$. 
The  volume element $\sqrt{\det(g_{AB})} = r^2(w, 0)\sqrt{q}$ with $\sqrt{q}$ being the volume element of a unit sphere metric, $q_{AB}$,  is nowhere zero on $\mathcal{B}^\epsilon(w)$ and for $\lambda=0$ on $\mathscr{N}^\epsilon(w)$. 
The function $r(w, 0)$ is extended for values $\lambda>0$ on every null hypersurface $w=$const to be an  areal distance so that the volume integral for a value $\lambda_c>0$ is given by $4\pi r^2(w,\lambda_c)$. 
A solution of the Einstein-Maxwell-scalar field equations on the domain $\mathcal{B}^\epsilon(w)\cup\mathscr{N}^\epsilon(w)$ can be found for  the double-null  CIBVP with the following specifications:
\begin{itemize}
    \item provide a value for $\epsilon$ to  specify the nature of the initial data surface $\mathscr{N}^\epsilon(0)$ with  $w=0$;
    \item specify the value of the gauge charge $q$;
    \item on $\mathscr{N}^\epsilon(0)$ specify a free (sufficiently differentiable) complex function $F_{\Phi}(\lambda)$;
    \item on $\mathcal{B}^\epsilon(0)$ specify a free (sufficiently differentiable) complex function $\mathcal{N}_{\mathcal{B}}(w)$;
    \item the following fields are trivial everywhere on $\mathcal{B}^\epsilon(w)$
    \begin{equation}
    \alpha(w,0)\;\;,\;\;V(w,0)\;\;,\;\;V_{,\lambda}(w,0).
    \end{equation}
    \item on the common intersection $\Sigma(0)=\mathcal{B}^\epsilon(0)\cap\mathscr{N}^\epsilon(0)$ specify the values for the following fields
    \begin{equation}\begin{split}
           &\Phi(0,0)\;,\;\;r(0,0)\;,\;\;r_{,w}(0,0)\;,\;\;\\
           &r_{,\lambda}(0,0)\;,\;\;Q(0,0).
           \end{split}
    \end{equation}
    where in particular $r(0,0)>0$.
\end{itemize}
This double-null CIBVP can be solved in the following manner
\begin{widetext}
\begin{enumerate}
\item With  $\mathcal{N}_{\mathcal{B}}(w)$ and $\Phi(0,0)$ determine the scalar field everywhere on $\mathcal{B}^ \epsilon(w)$ via the equation
    \begin{equation}
        \Phi_{,w}(w, 0) = \mathcal{N}_{\mathcal{B}}(w).
    \end{equation}
\item With the boundary data $\mathcal{N}_{\mathcal{B}}(w)$, the scalar field $\Phi(w,0)$ on $\mathcal{B}^\epsilon(w)$, and the initial values $r(0,0)$, $r_{,w}(0,0)$, $Q(0,0)$  and $r_{\lambda}(0,0)$ solve the hierarchical set of boundary equations
\begin{eqnarray}
0&=&
 r_{,ww}(w, 0)   
+\frac{1}{2}\kappa r (w, 0)
|\mathcal{N}_{\mathcal{B}}|^2 
   \;\;,\\
  0&=&
 Q_{,w} (w, 0) 
+2i\pi q r^2(w, 0)  \left[\overline {\Phi}(w, 0)\mathcal{N}_{\mathcal{B}} - \Phi(w, 0)   \overline{\mathcal{N}}_{\Phi} \right] \bigg|_{\mathcal{B}^ \epsilon}\;\;,\\
0&=&
 \left[
  2\epsilon (r r_{,\lambda})_{,w} 
 -1
+\frac{\kappa }{8\pi }\frac{Q^2}{r^2}
\right]\bigg|_{\mathcal{B}^\epsilon}.
\end{eqnarray}
 to obtain  the  fields $r$, $Q$ and $r_{,\lambda}$ on $\mathcal{B}^\epsilon(w)$. With the values of $r$, $r_{,w}$ and $\mathcal{N}_{\mathcal{B}}(w)$ on $\mathcal{B}^\epsilon$ determine the fields $Z$ and $\mathcal{L}$ everywhere on $\mathcal{B}^\epsilon(w)$. This determines all necessary  boundary values to integrate the hypersurface equations \eqref{eq:hyp_r}-\eqref{eq:hyp_V}.
\item On the initial null hypersurface $w=0$  utilize the initial value $\Phi(0,0)$ on $\Sigma(0)$ and the initial data $\mathcal{F}_{\Phi}$ to integrate 
\begin{eqnarray}
    \Phi_{,\lambda}(0, \lambda) = \mathcal{F}_{\Phi}(\lambda),
\end{eqnarray}
for the initial data $\Phi(0,\lambda)$.
\item on $\mathscr{N}^\epsilon(0)$, i.e. the null hypersurface $w=0$, integrate hierarchically the hypersurface equations \eqref{eq:hyp_r}-\eqref{eq:hyp_V} to obtain the fields $r(0, \lambda)$, $Q(0,\lambda)$, $\alpha(0,\lambda)$, $Z(0,\lambda)$, $\mathcal{L}(0,\lambda)$, and $V(0,\lambda)$, while using $\mathcal{F}_{\Phi}(\lambda)$, $\Phi(0,\lambda)$ and the respective boundary value fields evaluated at $w=0$.
\item With  $r(0, \lambda)$, $\mathcal{L}(0,\lambda)$, $V(0,\lambda)$, and $\Phi_{,\lambda}(0,\lambda)$ algebraically calculate $\Phi_{,w}(0,\lambda)$ on $w=0$ using  Eq. \eqref{ev_eqn_Phi}, integrate this $w$ derivative (using, for example, a finite difference scheme) to find $\Phi(\Delta w, \lambda)$ at some time $w=\Delta w\neq 0$ on a null hypersurface $\mathscr{N}^\epsilon(\Delta w)$
    \item The new values $\Phi(\Delta w, \lambda)$ serve as initial data on $\mathscr{N}^\epsilon(\Delta w)$ and are used to determine the metric and electromagnetic fields following step 4 above but using the coordinate time $w=\Delta w$ instead of $w=0$.
\end{enumerate}
\end{widetext}
Comparing this formulation of the double-null CIBVP with the one of the previous section Sec.~\ref{sec:dn_req0}, reveals that the two are very similar. In fact, the only essential difference is step 2 -- the determination of the nontrivial boundary fields in $\mathcal{B}^\epsilon(w)$. 
The reason for this difference is that for one of the formulations $r=0$ on $\mathcal{B}^\epsilon(w)$, while for the other one it is essentially nontrivial. 

The two different formulations give rise for different areas of application. 
The first scheme, where $r=0$ on $\mathcal{B}^\epsilon(w)$, is well suited for studying the evolution of  initial data with a regular center. 
The second scheme is suited for the study of (charged) black holes in which $\mathcal{B}^\epsilon(w)$ is a horizon with an associated horizon mass of the black hole. Indeed, in Sec.~\ref{sec:RN_null} the second scheme is used to derive the  Reissner-Nordstr\"om solution of a charged black hole.

\section{Symmetry considerations}\label{sec:symmetry}
For the subsequent sections, we consider possible timelike Killing vectors $\xi$. Killing equations are given by 
\begin{equation}
    \label{eq:KE}0=\nabla^a\xi^b+\nabla^b\xi^a = -g^{ab}_{\;\;,c}\xi^c
    +g^{cb}\xi^a_{\;\;,c}
    +g^{ac}\xi^b_{\;\;,c}\;.
\end{equation}
An ansatz for the Killing vector $\xi  $ left invariant under  spatial rotations is  $\xi  = \xi^w\partial_w+\xi^\lambda\partial_\lambda$ with the components $\xi^w$ and $\xi^\lambda$ only dependent on $w$ and $\lambda$. Using \eqref{eq:inverse_g}  we find 
\begin{subequations}
    \begin{align}
    0=& 2\epsilon \xi^w_{\;,\lambda}\;,\\
    0=& \epsilon \xi^w_{\;,w}+\epsilon \xi^\lambda_{\;,\lambda}
    +V\xi^w_{,\lambda}\;,\\
    0=& -\frac{2}{r^3}\left(r_{ ,w}\xi^w+r_{ ,\lambda}\xi^\lambda\right)\label{eq:KE_r}\;,\\
    0=& -V_{,w}\xi^w-V_{,\lambda}\xi^\lambda
    +2\epsilon \xi^\lambda_{\;,w}
    +2V \xi^\lambda_{\;,\lambda}\;.\label{eq:KE_V}
\end{align}
\end{subequations}
From the first two, we find
\begin{equation}\label{sol_KV}
\xi  = A(w) \partial_w + \left[-A_{,w}(w)\lambda +B(w) \right]\partial_\lambda\;,
\end{equation}
where $A$ and $B$ are free functions. 
It is convenient to introduce  the scalar  function
\begin{equation}
\label{eq:z}
z = A(w)\lambda - \int^w B(\hat w)d\hat w\;,
\end{equation}
which is invariant under the action of the vector field $\xi$ that is the Lie derivative $\mathscr{L}_\xi z = \xi^a\partial_az=0$.
In principle, we may choose $B=0$, because of the metric's invariance of the coordinate transformation of the type \eqref{eq:invariance_metric}. However, we will keep it general to allow a broader family of affine-null coordinates. 
Inserting \eqref{sol_KV} into \eqref{eq:KE_r} yields
\begin{equation}
    0=\frac{2}{r^3}\left[r_{ ,w}A-(\lambda A_{,w}-B)r_{ ,\lambda}\right]\;,\label{eq:KE_rr}
\end{equation}
whose general solution  can be found  as
\begin{equation}\label{eq:r_z}
    r = \hat r(z)\;,
\end{equation}
where $\hat r(\cdot)$ is any nonvanishing differentiable function.
Finally,  inserting \eqref{sol_KV} into \eqref{eq:KE_V} gives us
\begin{equation}
\begin{split}
    0=&
    V_{,w}A-V_{,\lambda}(A_{,w}\lambda-B)
    -2\epsilon\left( A_{,ww}\lambda -B_{,w}\right)
    +2VA_{,w}
\end{split}
\end{equation}
with the general solution 
\begin{equation}\label{eq:Vgeneral}
V =\frac{\hat V[z(w, \lambda)]}{A^2(w)} 
-\frac{2\epsilon(A_{,w}\lambda-B)}{A}\;,
\end{equation}
in which $\hat V(\cdot)$ is an arbitrary differentiable function.

\begin{widetext}
Some choices for $A$ and $B$ relevant in this work are listed in Table ~\ref{Tab:killing}.
\begin{table}[h!]
    \centering  
     \begin{equation*}
    \begin{array}{|c|c||c|c|c|c|}\hline
       \quad A\quad &\quad B\quad &\quad\xi\quad&\quad \mbox{Invariant}\;\;z&\qquad r\qquad&\qquad V\qquad  \\[0.5em]\hline\hline
       A(w)&B(w)&A(w) \partial_w - \left[A_{,w}(w)\lambda -B(w) \right]\partial_\lambda&A(w)\lambda - \int^w B(\hat w)d\hat w&\hat r(z)&  
       \frac{\hat V(z)-2\epsilon A(w)[A_{,w}(w)\lambda-B(w)]}{A^2}
       \\[0.5em]\hline 
       1&0&\partial_w&\lambda&\hat r(z)&  \hat V(z)\\[0.5em]\hline 
       kw&0&k(w\partial_w-\lambda\partial_\lambda)& kw\lambda&\hat r(z)&
       \lambda^2\frac{\hat V(z)-2\epsilon kz}{z^2}=\lambda^2 \tilde{V}(z)\\[0.5em]\hline
       \frac{1}{2}\left(\frac{w}{w+k}\right)^2
       &1
       &
       \frac{1}{2}\left(\frac{w}{w+k}\right)^2\partial_w + \left[-\frac{kw\lambda}{(w+k)^3} +1\right]\partial_\lambda
       &
       \frac{1}{2}\left(\frac{w}{w+k}\right)^2\lambda-w
       &
       \hat r(z)
       & \frac{4(w+k)^4\hat V\left(z\right)}{w^4}
       +\frac{4\epsilon}{w^2}\left[(w+k)^2 - \frac{ kw\lambda}{w+k}\right]\\[0.5em]\hline
    \end{array}
    \end{equation*}
    \caption{Selective choices for the $A$ and $B$ to build a Killing vector.}
    \label{Tab:killing}
\end{table}
\end{widetext}

The coordinate dependence of the metric potentials $r$ and $V$ for the different expressions for  Killing vectors  in Table \ref{Tab:killing} is general. 
For further particular applications, it is practical to study the norm of the Killing vector either for large values of the affine parameter or at its origin. 
The former will give us restrictions to the metric by requiring the Killing vector be  asymptotically timelike, while the latter gives us necessary conditions, if the origin of the affine parameter is on a Killing horizon with a given surface gravity.   

The norm of the Killing vector \eqref{sol_KV} is
\begin{equation}\label{norm_gKV_}
    \xi^a\xi_a = -VA^2-\epsilon (A^2)_{,w}\lambda  +2\epsilon AB.
\end{equation}
In its asymptotic limit ($\lambda\rightarrow\infty$), we find using \eqref{eq:V_inf}, \eqref{eq:Uinf} and \eqref{eq:Vinf}: 
\begin{align}\label{eq:KV_asympt}
    \xi^a\xi_a
    =&-2\epsilon A^2\left[ \left(\ln \frac{A}{H}\right)_{,w}
    \right]\lambda\nonumber\\
    &+
    \left\{2\epsilon B -\frac{[1 
    -2\epsilon H(R_\infty)_{,w}] A}{H^2}\right\}A
    +O(\lambda^{-1}).
\end{align}
Hence, as seen in \eqref{eq:KV_asympt}, the asymptotic behavior of the Killing vector norm is determined by the asymptotic behavior of the areal distance through the functions $H$ and $R_\infty$.
Therefore,  in order to have a finite norm of this asymptotic limit, the linear term in $\lambda$ should vanish. 
One possibility is to set $A=0$, which from \eqref{norm_gKV_} implies the Killing vector field $\xi=B\partial_\lambda$ with vanishing norm. However, taking into account \eqref{eq:r_z}, we see that it implies that the vector field is zero everywhere or $\xi$ is a null Killing vector with $r=r(w)$ everywhere, which is inadmissible.

The other choice for the finiteness of the norm of the Killing vector for $\lambda\rightarrow \infty$ is 
\begin{equation}\label{eq:AcH}
   \left(\ln \frac{A}{H}\right)_{,w} = 0\;,
\end{equation}
whose integration immediately gives
\begin{equation}\label{sol_KV_A}
    A
    =
    cH\;,
\end{equation}
where $c$ is a nontrivial constant.

If the Killing vector is required to be an asymptotically timelike unit  vector,  one finds the condition
\begin{equation}\label{eqref:Bfinal}
    -1 =  \left[2\epsilon B -\frac{c 
    -2c\epsilon H(R_\infty)_{,w}}{H}\right]cH,
\end{equation}
or
\begin{equation}\label{sol_KV_B}
B =\frac{c \epsilon}{2 H} -  c(R_\infty)_{,w} -\frac{\epsilon}{2 cH}.
\end{equation}
The relations \eqref{sol_KV_A} and \eqref{sol_KV_B} show that the components of a timelike Killing vector  can be completely fixed by the asymptotic solution of the field equation.
In particular, 
it is determined by the asymptotic behavior of the areal distance $r$. An asymptotic timelike Killing vector should therefore have the form
\begin{equation}\label{eq:KV_general_sol}
\begin{split}
    \xi =& cH(w)\partial_w \\
    &+\left[-cH_{,w}\lambda +\frac{c \epsilon}{2 H} -  c(R_\infty)_{,w} -\frac{\epsilon}{2 cH}\right]\partial_\lambda
    +O(\lambda^{-1}).
\end{split}
\end{equation}

In particular, regarding the choices of $A$ and $B$ in Table \ref{Tab:killing} and requiring the Killing vectors to be asymptotically  timelike  unit  vectors implies using \eqref{eq:r_z}, \eqref{eq:AcH} and \eqref{sol_KV_B}: 
\begin{enumerate}
\renewcommand{\labelenumi}{\roman{enumi})}
    \item For case $A=1$ and $B=0$, we find $c=1/H$. Moreover, since $r=r(\lambda)$,  $H$=const and $R_\infty$=const. Consequently, $\xi^a\xi_a=-1$ in the asymptotic limit can only be achieved if $H=1$; i.e. the frame at large radii is a Bondi frame and the areal distance has the asymptotic expansion
    \begin{equation}
        r  =  \lambda +R_\infty+O(\lambda^{-1}).
    \end{equation}
    \item The case $A=kw$ and $B=0$ gives $kw  = cH(w)$.  As Table \ref{Tab:killing} 
 shows $r=r(w\lambda)$,   the asymptotic expansion of $r$ in \eqref{eq:r_inf} becomes  $r(w\lambda) = \frac{kw\lambda}{c}+R_\infty(w)+O(\lambda^{-1})$, which requires $R_\infty$=const. 
 But then the normalization $\xi^a\xi_a = -1$ implies $c^2=1$, so that we can choose $c=1$. Consequently $r$ has the asymptotic expansion 
    \begin{equation}
        r  =  kw\lambda +R_\infty+O(\lambda^{-1}).
    \end{equation}
    \item The case $A=\frac{1}{2}w^2(w+k)^{-2}$ and $B=1$ gives ${w^2 = 2 c(w+k)^2H}$. Then fixing $H$ in  the asymptotic behavior of $r$ gives for \eqref{eq:r_inf} 
    \begin{equation}\label{eq:r_KV_case_iii}
        r(z) = \frac{w^2 c}{2(w+k)^2}\lambda + R_\infty(w)+O(\lambda^{-1})\;\;.
    \end{equation}
    Table \ref{Tab:killing} indicates that
    $$r(z) = r\left[\frac{1}{2}\left(\frac{w}{w+k}\right)^2\lambda-w\right],$$
    which implies by comparison with \eqref{eq:r_KV_case_iii} that
    \begin{equation}
    R_\infty(w) = -c w+r_\infty\;\;, r_\infty=\mathrm{const}.
    \end{equation}
    To be consistent with the invariant,  any solution for $r$ of the field equations  with this Killing vector, which is in addition an asymptotical timelike unit vector, must have the large $\lambda$ expansion  of the form
    \begin{equation} 
        r(z) =r_\infty +c\left[ w- \frac{w^2 }{2(w+k)^2}\lambda\right] 
        + O(\lambda^{-1}).
    \end{equation}
    The determination of $c$ for the present case is postponed to Sec.~\ref{sec:RN_null}, because of the subtlety of the behavior of the  Killing vector for $w=0$. 
    There, it is a nonvanishing null tangent vector to the hypersurface, consequently, $\xi|_{w=0}=\partial_\lambda$. 
     {A regularity requirement of this surface [which implies regularity of $V$ given by \eqref{eq:Vgeneral}]} can be used to fix the value of the constant $c$ (see next section for an explicit example in an extremal Reissner-Nordstr\"om black hole).
\end{enumerate}

Of interest is also the behavior of the norm of the Killing vector for $\lambda=0$, i.e on $\mathcal{B}^\epsilon$. At the origin ($\lambda=0$) of the affine parameter, the norm \eqref{norm_gKV_} takes the value 
\begin{equation}\label{norm_gKV}
    \xi^a\xi_a|_{\mathcal{B}^\epsilon} = -A^2V|_{\mathcal{B}^\epsilon} +2\epsilon AB.
\end{equation}
If $V|_{\lambda=0}=0$  the surface $\lambda=0$ is a null hypersurface $\mathcal{B}^\epsilon$, and 
if  $B=0$ and $A\neq 0$ the null hypersurface $\mathcal{B}^\epsilon$ is a Killing horizon\footnote{{If $A=0$, we also have $\xi^a\xi_a=0$, but it is not a Killing horizon because only a vector field of the form $\xi\propto\partial_w$ can be a null generator of a Killing horizon at a null surface determined by $\lambda=0$. 
However, the possibility $A=B=0$ can be satisfied at an eventual bifurcating Killing horizon, where the Killing vector field vanishes}.}.  Indeed, $B\neq0$  in case (iii) is crucial for  finding the (vanishing) surface gravity of an extremal black hole   as is the case of extremal Reissner-Nordstr\"om black holes discussed in the next section. It turns out that for case iii) there also exists a Killing horizon  not necessarily located at $\lambda=0$ but given implicitly by an expression of the form $F(w,\lambda)=0$.  In the other cases, 
the corresponding surface gravity, $\kappa_H$,  can be found from the invariant relation \cite{1984ucp..book.....W}
\begin{equation}\label{eq:surface_gravity}
(\kappa_H)^ 2  = - \frac{1}{2}(\nabla^a\xi^b)(\nabla_a\xi_b)\Big|_{\mathcal{B}^\epsilon} 
= \left(A_{,w}   
  - \frac{A}{2}V_{,\lambda}|_{\mathcal{B}^\epsilon}\right)^2 .
\end{equation}
These properties of the Killing vector will be used in the next section, where the Reissner-Nordstr\"om black hole is discussed.

\section{Exact solutions for the ciBVP}
\subsection{Charged black hole solutions in null coordinates}\label{sec:RN_null}
For finding charged black hole solutions, we work with the double-null foliation of Sec.~\ref{sec:cibvp_dn}, where the null boundary has nontrivial  Misner-Sharp mass.

We assume that the scalar field  on the initial slice $\mathscr{N}^\epsilon(0)$ and the boundary $\mathcal{B}^\epsilon(w)$ is trivial meaning $\mathcal{F}_\Phi=\mathcal{N}_{\mathcal{B}}  = 0$.
  We use $\kappa=8\pi$ and the most general values for the remaining data at the common intersection $\Sigma(0)$
\begin{equation}
\begin{split}
&    r(0,0)=r_0>0\;\;,\;\;
r_{,w}(0,0)=r_N\;\;,\;\;\\
&Q(0,0)=Q_0\;\;,\;\;
r_{,\lambda}(0,0) = \Theta.
\end{split}
\end{equation}
The boundary equations on $\mathcal{B}^\epsilon$ are 
\begin{subequations}\label{bound_H_phi_eq0}
\begin{eqnarray}
\frac{r_{,ww}}{r}\bigg|_{\mathcal{B}^\epsilon} &=&
0,\label{bnd_RN_r}
\\
 Q_{,w}\big|_{\mathcal{B}^\epsilon}&=&
 0,\label{bnd_RN_Q}
\\
(rr_{,\lambda})_{,w} \big|_{\mathcal{B}^\epsilon}
&=&
\frac{\epsilon}{2}\left( 
 1
- \frac{ Q^2}{  r^2} \right)\bigg|_{\mathcal{B}^\epsilon}.
\label{bnd_RN_rlambda}
\end{eqnarray}
\end{subequations}
Integration of \eqref{bnd_RN_r} and \eqref{bnd_RN_Q}  gives us
\begin{subequations}\label{sol_r_Q_RN}
    \begin{align}
 r(w , 0) =& r_{\mathcal{B}}:=r_0+r_Nw  \;\;,\\
Q(w , 0) =& Q_0\;\;.\;\;
\end{align}
\end{subequations}
Insert these values into \eqref{bnd_RN_rlambda} and integration yields

\begin{equation}\label{sol_r1}
\begin{split}
(r_{,\lambda})  \big|_{\mathcal{B}^\epsilon}
 =
 &  r_{1}(w ):= 
  \frac{r_0\Theta}{r_{\mathcal{B}}}
 + \frac{ \epsilon}{2 }\left(\frac{r_0r_{\mathcal{B}}
-Q_0^2  }{r_0r_{\mathcal{B}}^ 2} \right)\left(\frac{r_{\mathcal{B}}-r_0}{r_N }\right)\\
=&\frac{\Theta r_0}{r_Nw +r_0} +\frac{(r_0r_Nw-Q_0^2+r_0^2)\epsilon w}{2r_0(r_Nw +r_0)^2}
\;\;.
\end{split}
\end{equation}
The boundary values \eqref{sol_r_Q_RN} and \eqref{sol_r1} provide the values for $Z$ and $\mathcal{L}$ at $\mathcal{B}^\epsilon$ 
\begin{equation}\label{eq:bnd_YL_RN}
Z(w , 0)  = 2\epsilon r_{\mathcal{B}} r_N\;\;,\;\; \mathcal{L}(w, 0) = 0\;\;.
\end{equation}
Now the hypersurface  equations for the given initial boundary value data are
\begin{subequations}\label{hyp_ev_gRN}
\begin{eqnarray}
 r_{,\lambda\lambda}
&=& 
0 ,\\
 Q_{,\lambda} 
&=& 0,\\
\alpha_{,\lambda}&=&-\epsilon\frac{Q}{r^2} , \\
Z_{,\lambda} 
&=&
 -\frac{Q^2}{r^2}  , \\
\mathcal{L}_{,\lambda}&=&
0,\\
V_{,\lambda\lambda} &=&
-\frac{1}{\lambda}\left(\frac{\lambda^2}{r^2}{}\right)_{,\lambda}
 +\frac{2 Zr_{,\lambda}}{r^3}
 +\frac{4Q^2 }{r^4}.
 \end{eqnarray}
\end{subequations}
The integration of \eqref{hyp_ev_gRN} with the boundary data \eqref{sol_r_Q_RN}, \eqref{sol_r1}, and \eqref{eq:bnd_YL_RN} gives us
\begin{subequations}\label{eq:gen_sol_RN}
\begin{eqnarray}
r  &=& r_{\mathcal{B}}+r_1 \lambda ,\label{eq:rwrhlambda}\\
Q  &=& Q_0,\\
\alpha &=&\epsilon \frac{Q_0}{ r_{1}}\left(\frac{1}{r} -\frac{1}{r_{\mathcal{B}}}\right), 
    \\
Z 
&=& 
2\epsilon r_{\mathcal{B}} r_N
+ \frac{Q_0^ 2}{r_{1}}\left(\frac{1}{r} -\frac{1}{r_{\mathcal{B}}}\right), \\
V 
&=&
\left(\frac{\lambda}{r}\right)^ 2\frac{Q^2_0(r+r_\mathcal{B}) -(1-2\epsilon r_N r_1)rr_\mathcal{B}^ 2  }{r_\mathcal{B}^ 3}\nonumber\\
&=&
\left(\frac{\lambda}{r}\right)^ 2
\left[\frac{Q_0^2}{r_\mathcal{B}^2}+
\frac{\left(Q^2_0 - r_\mathcal{B}^2+2\epsilon r_N r_1r_\mathcal{B}^ 2\right)r}{r_\mathcal{B}^3}
\right].
\end{eqnarray}
\end{subequations}
Consequently, the most general form of {the metric for these initial values reads}
    \begin{equation}
        \begin{split}
            ds^2
            =& 
            -\left(\frac{\lambda}{r}\right)^ 2
\left[\frac{Q_0^2}{r_\mathcal{B}^2}+
\left(Q^2_0 - r_\mathcal{B}^2+2\epsilon r_N r_1r_\mathcal{B}^ 2\right)\left(\frac{r}{r_\mathcal{B}^3}\right)
\right]dw^2  \\
&
            + 2\epsilon dwd\lambda + r^2q_{AB}dx^Adx^B\;,\\
            r=&r_0+r_N w + \left[\frac{\Theta r_0}{r_Nw +r_0} +\frac{(r_0r_Nw-Q_0^2+r_0^2)\epsilon w}{2r_0(r_Nw +r_0)^2}\right]\lambda,
        \end{split}
    \end{equation}
in which $Q_0$ corresponds to a global charge. This solution was also found in \cite{Gallo:2021jxt} using a different approach.

On $\mathcal{B}^\epsilon$, it reduces to
\begin{equation}
            ds^2
            = (r_0+r_N w)^2q_{AB}dx^Adx^B,
    \end{equation}
and on $\mathscr{N}^\epsilon(0)$, where  $w=0=dw$, 
\begin{equation}
            ds^2
            = (r_0+\Theta \lambda)^2q_{AB}dx^Adx^B.
\end{equation}
Since $w$  and $\lambda$ are the affine parameters on the null hypersurfaces $\mathcal{B}^\epsilon(w)$ and $\mathscr{N}^\epsilon(0)$, respectively, the initial data parameters $r_N$ and $\Theta$ correspond to the expansion rates of the generators of these surfaces. 

Calculation of the Misner-Sharp mass \eqref{eq:misner_sharp_mass} gives us
\begin{align}\label{eq:MSM_bh_general}
    M
    =&\frac{1}{2}\left[ 
    r_\mathcal{B}   
    - 2\epsilon r_\mathcal{B} r_N r_1
    +Q^ 2_0\left(\frac{1}{r_\mathcal{B}}-\frac{1}{r}\right)\right],   
\end{align}
with the limiting values 
\begin{align}
    \lim_{w\rightarrow0} M =&
    \frac{1}{2}\left[ 
    r_0   
    - 2\epsilon r_0 r_N \Theta
    +Q^ 2_0\left(\frac{1}{r_0}-\frac{1}{r_0+\Theta \lambda}\right)\right],
\\
    \lim_{\lambda\rightarrow0} M =&
    \frac{1}{2}(r_0+r_N w )\left( 
      1
    - 2\epsilon  r_N r_1
     \right) ,
\end{align}
so that the composite limit
\begin{align}
    \lim_{w\rightarrow0 \atop \lambda\rightarrow0 } M =&
    \frac{r_0}{2}(
    1   
    - 2\epsilon  r_N \Theta),
\end{align}
shows that $r_0=2M$ if either $r_N$ or $\Theta$ vanishes.

As the limit 
\begin{align}
\lim_{\lambda\rightarrow \infty}M =& \frac{r_\mathcal{B} }{2}\left[ 
     1
    - 2\epsilon   r_N r_1
    + \frac{Q^ 2_0}{r_\mathcal{B}^ 2} \right],  
\end{align}
gives rise to the Bondi mass, we find the most general expression for the Bondi mass of the solution to be
\begin{equation}\label{eq:mB_general}
    m_B = \frac{Q_0^2+r_0^2  }{2r_0} -\epsilon r_0 r_N\Theta\;.
\end{equation}

If either $r_N=0$ or $\Theta=0$,  the initial value $r_0$  can be fixed by the Bondi mass, $m_B$, to find 
\begin{equation}\label{eq:RN_horizons}
r_{0,\pm} = m_B\pm\sqrt{m_B^ 2 - Q^2_0},
\end{equation}
where the numerical value $r_{0,+}$ is the same value as the value of the radius of the outer event horizon of a Reissner-Nordstr\"om black hole and $r_{0,-}$ the value of the inner event horizon.

\subsubsection{Subextremal black holes}

Suppose $r_N=0$; then the areal distance has the  form

\begin{equation}
    r = r_0 +\left[\Theta -\epsilon\frac{Q^2_0-r^2 _0}{2r_0^ 3} w  \right]\lambda\;\;.
\end{equation}
Since $w$ parametrizes affine geodesics on the boundary $\mathcal{B}^\epsilon$ where $\lambda=0$, we have the freedom of choice in the origin of the affine parameter. The transformation 
\begin{align}
    w\rightarrow \tilde{w}= w-\epsilon \Theta \frac{2 r_0^3}{Q_0^2-r_0^2},
\end{align}
gives the symmetric expression 
\begin{subequations}\label{eq:non_extreme}
\begin{align}\label{eq:r_subextremal}
r=\tilde r(\lambda \tilde w) := r_0 -\epsilon\frac{Q^2_0-r^2 _0}{2r_0^ 3} (\tilde{w}  \lambda),
\end{align}
and 
\begin{align}
V &= -2\lambda^ 2\frac{\epsilon (\tilde{w} \lambda)(Q_0^2-r_0^2)^2 -2 r_0^ 4(2Q^2_0-r_0^ 2)}{\left[\epsilon (\tilde{w} \lambda)(Q_0^2-r_0^2)
-2r_0^4\right]^2}\nonumber\\
&\equiv \lambda^2 \tilde V(\tilde{w}\lambda),
\end{align}
\end{subequations}
thereby showing the two functions depend on the product $w\lambda$. 
For simplicity, we will write $\tilde{w}$ as $w$, hereafter.

Therefore, the metric 
\begin{equation}\label{eq:metr-non-extr}
    \begin{split}
        ds^2 = -\lambda^2\tilde V(\lambda w)   dw^2+2 \epsilon dwd\lambda + \tilde r ^2(\lambda w)q_{AB}dx^Adx^B,
    \end{split}
\end{equation}
is invariant under the transformation $(w, \lambda)\rightarrow( a w, a^{-1}\lambda)$ for any number $a$, which implies a Killing vector of the form
$\xi = \epsilon k(w\partial_w  - \lambda\partial_\lambda)$, where $k$ is a positive normalization constant determined from the calculation of the norm $\xi^ a\xi_a$ and the factor $\epsilon$ is included in order to choose a future timelike Killing vector field. In accordance with the results of Sec.~\ref{sol_KV}, this also gives us $A(w) = kw$ and $B=0$ for the two $w-$dependent functions of the general Killing vector in Eq. \eqref{sol_KV}.
Calculation of the asymptotic limits of the norms gives us
\begin{align}
\lim_{\lambda\rightarrow 0 }\xi^ a\xi_a=&0\;\;,\\
\lim_{\lambda\rightarrow \infty }\xi^ a\xi_a=& -\left[\frac{2k r_0^ 3}{Q_0^2-r_0^ 2 }\right]^2\;\;.
\end{align}

Therefore, the Killing vector is an asymptotic  timelike unit  vector if the Killing vector has the form
\begin{equation}\label{eq:KV_nonextremal}
    \xi = 
    \frac{\epsilon}{2r_0}
    \left(1-\frac{Q^2_0}{r_0^ 2}\right)(w\partial_w  - \lambda\partial_\lambda)\;\;,
\end{equation}

As mentioned above, if the norm of the Killing vector $\xi^a\xi_a$ vanishes for $\lambda=0$, the null hypersurface $\mathcal{B}^\epsilon$ is a Killing horizon. If $\epsilon=1$, this boundary is the future Killing horizon $H^+$, and the past event horizon $H^-$ is for the choice $\epsilon=-1$. 
Its surface gravity $\kappa_H$ can be found from the invariant relation \cite{1984ucp..book.....W}
\begin{equation}
\kappa_H^ 2  
= - \frac{1}{2}(\nabla^a\xi^b)(\nabla_a\xi_b)\Big|_{ \mathcal{B}^\epsilon } 
= k^ 2 
=\left[\frac{1}{2r_0}\left(1-\frac{Q^2_0}{r_0^ 2}\right)\right]^ 2\;\;.
\end{equation}
Using the value of $r_{0,+}$ of \eqref{eq:RN_horizons}, the surface gravity reduces to the value of the Reissner-Nordstr\"om black hole
\begin{equation}\label{eq:sGrav_subextremal}
  \kappa_H = \frac{\sqrt{m^2_B-Q_0^ 2}}{2mr_{0,+} - Q_0^ 2} = \frac{\sqrt{m^2_B-Q_0^ 2}}{2m_B\left(m_B+\sqrt{m^ 2_B-Q_0^ 2}\right)  - Q_0^ 2}\;,  
\end{equation}
which also gives the Schwarzschild value ${\kappa_H = 1/(4m_B)}$ 
 in the case of vanishing global charge, $Q_0=0$.

 Hence, the Killing vector field can be written as
 \begin{equation}
     \xi=\epsilon\kappa_H(w\partial_w-\lambda\partial_\lambda).\label{eq:killingsurfgrav}
 \end{equation} Note also that the metric \eqref{eq:metr-non-extr} is regular at the null surface $w=0$ given by $\mathscr{N}^\epsilon(0)$. 
 This surface is the future (past) Killing horizon $H^+$ ($H^-$) if $\epsilon=-1$ ($\epsilon=1$)  with Killing vector field $\xi=\kappa_H\lambda\partial_\lambda$ ($\xi=\kappa_H w\partial_w$). On the other hand, $\xi$ vanishes at the two-sphere  given by $w=\lambda=0$, which is a bifurcating Killing horizon.

\subsubsection{Extremal black holes}

The surface gravity $\kappa_H$ in \eqref{eq:sGrav_subextremal} 
vanishes, if $Q_0=m_B=r_0$, but this goes along with the vanishing of the Killing vector in \eqref{eq:KV_nonextremal}. 
In fact, in that limit the resulting metric associated with the $\{w,\lambda\}$ coordinates introduced in the last subsection is not the extremal Reissner-Nordstr\"om metric, but the so-called Bertotti-Robinson metric \cite{Gallo:2021jxt}. Thus the Killing vector \eqref{eq:KV_nonextremal} cannot be used to find an extremal  Reissner-Nordstr\"om  black hole solution where the global charge-to-mass ratio is unity.  Indeed, as we will show here, the resulting extremal metric is not invariant under the transformation $(w, \lambda)\rightarrow( a w, a^{-1}\lambda)$, and therefore the resulting Killing vector field cannot be written as in \eqref{eq:KV_nonextremal}.

To single out an extremal black hole solution in the general solution \eqref{eq:gen_sol_RN}  for $r$ and $V$, which also has a nonvanishing Killing vector at the horizon of the black hole, consider \eqref{eq:mB_general}, while setting $m_{e}:=Q_0=m_B.$
This gives
\begin{equation}
    m_{e} = \frac{m_{e}^2+r_0^2  }{2r_0} -\epsilon r_0 r_N\Theta,
\end{equation}
allowing us to express any two of the initial data $(r_0,\,r_N,\Theta)$, by the third, i.e.
\begin{align}
    r_0=& \frac{m_{e}}{2\epsilon r_N\Theta-1}\left(-1\pm\sqrt{2\epsilon r_N\Theta}\right)\label{eq:r0_by_rnTheta}\;,\\
    r_N=&\frac{(m_{e}-r_0)^2}{2\epsilon r_0^2\Theta}\label{eq:rN_by_r0Theta}\;,\\
    \Theta=&\frac{(m_{e}-r_0)^2}{2\epsilon r_0^2r_N}\;.\label{eq:Theta_by_r0rN}
\end{align}
Since in the Schwarzschild case $r_0$ equals twice the Bondi mass, it is desirable to relate $r_0$ one-to-one with $m_e$. In \eqref{eq:r0_by_rnTheta} this is achieved only by setting either  $\Theta=0$ or $r_N=0$ giving in both cases $r_0=m_e$. 
In \eqref{eq:rN_by_r0Theta}, the choice $r_0=m_e$ gives $r_N=0$ and $\Theta\neq0$ left arbitrary, while for \eqref{eq:Theta_by_r0rN} the same choice yields $\Theta=0$ and $r_N\neq0$ left arbitrary.

Thus an extremal black hole solution can be found if $r_0=m_e=Q_0$ and either $r_N=0$ and $\Theta\neq 0$ or $\Theta=0$ and $r_N\neq0$.
Now consider the solution  for the areal distance and metric potential $V$  {for these  cases:} for the case $(r_N = 0\;,\;\Theta\neq0)$ we find
\begin{subequations}
\begin{align}
    r =& m_e + \Theta \lambda\label{eq:r_RN_extremal_Bondi}\\
    V =& \left(\frac{\lambda}{m_e+\Theta\lambda}\right)^2,
\end{align}
\end{subequations}
and for the case $(\Theta = 0\;,\;r_N\neq0)$
\begin{widetext}
    \begin{align}
    r=& m_e + r_Nw +\left[\frac{\epsilon r_Nw^2}{2(r_Nw +m_e)^2} 
   \right]\lambda \;,\label{eq:rextrem}\\
   V=&
   {-\frac{4m_e\lambda^2
   [\epsilon\lambda r_N^2 w^3 - (m_e-2r_Nw )(m_e+r_Nw )^3]}{(r_Nw +m_e)[w^4 r_N^2\lambda^2 + 4\epsilon r_N w^2(m_e+r_Nnw  )^3\lambda + 4(m_e+r_N w  )^6]}}\;.\label{eq:vextrem}
\end{align}
\end{widetext}
The line element of the first solution is 
\begin{equation}\label{eq:RNextrmal_KV1}
\begin{split}
    ds^2 =& -\left(\frac{\lambda}{m_e+\Theta\lambda}\right)^2d w^2
    +2\epsilon d w d\lambda \\
    &+(m_e+\Theta\lambda)^2q_{AB}dx^Adx^B\;,
\end{split}
\end{equation}
but, in fact, it looks more familiar if $r(\lambda)$ in \eqref{eq:r_RN_extremal_Bondi} is inverted to $\lambda(r)$, and  then we find 
\begin{equation}
V(r) = \frac{1}{\Theta^2}\left(1-\frac{m_e}{r}\right)^2\;\;,\;\;
\end{equation}
so that the coordinate transformation $\hat w = w/\Theta$, $\lambda = (r-m_e)/\Theta$ leaves us with the well-known line element
\begin{equation}
ds^2 = -\left(1-\frac{m_e}{r}\right)^2 d\hat w^2+2\epsilon d\hat w dr +r^2q_{AB}dx^Adx^B.
\end{equation}
of an extremal Reissner-Nordstr\"om black hole with either ingoing ($\epsilon=1$) or outgoing ($\epsilon=-1$) polar null coordinates. 
In the first case, $\hat{w}$ is usually represented as $v$, and the past event horizon is reached as $v\to -\infty.$ 
For $\epsilon=1$, this metric covers the interior region of the past event horizon and the exterior region  (in that case $\hat{w}$ is identified with the retarded null coordinate $u$). This coordinate does not cover the future horizon, which is asymptotically reached as $u\to\infty$. 
Indeed, calculation of the outgoing expansion rate of the surface forming null vector  $\ell^a\partial_a=\partial_\lambda$ using \eqref{eq:RNextrmal_KV1} gives $2\Theta/(m_e+\Theta\lambda)$ which does not vanish for any value of  $w$. 
Thus there is no  null surface {$w=$const}  labeling the future event horizon. 
In a similar way, for ingoing null coordinates ($\epsilon=1$), the resulting coordinates cover the exterior and interior of the future event horizon.  

Thus the solution with initial data $r_N=0$ and $\Theta\neq0$ cannot describe an extremal black hole in a regular manner, i.e. including both horizons. 
However, we will see that the description of an extremal charged black hole with both horizons is possible for  initial data with $\Theta=0$ and $r_N\neq0$.

Consider now the case $\Theta=0$, which gives \eqref{eq:rextrem} and \eqref{eq:vextrem} for the non trivial components of the metric.

 From \eqref{eq:rextrem}  and \eqref{eq:r_inf}, we read off the relevant functions, $H$ and $R_\infty$ of the asymptotic expansion of $r$ as
\begin{eqnarray}\label{eq:H_Ring_BH_extr}
        H(w)&=&\frac{\epsilon r_Nw^2}{2(r_Nw +m_e)^2} \;,
     \\
    R_\infty(w)&=& m_e + r_Nw\;\;.
\end{eqnarray}
Also note  the surface $w=0$ is a regular null surface determined by $r=m_e$ and where $V=\frac{\lambda^2}{m^2_e}$.
From  \eqref{eq:H_Ring_BH_extr} and \eqref{eq:KV_general_sol}, we find the following asymptotic behavior of the Killing vector near $w=0$

\begin{equation}
    \xi = \left[\frac{c^2-1}{c r_N w^2}{(2r_Nw+m_e)m_e-\frac{r_N}{c}}\right]\partial_\lambda + O(w)\;,
\end{equation}
which implies $c=\pm 1$ for the Killing vector be  regular   and finite at $w=0$. Setting $c=1$, we obtain from \eqref{eqref:Bfinal} that $B=-r_N.$ Therefore, from \eqref{sol_KV} and \eqref{eq:AcH} we arrive to the final expression for the  Killing vector $\xi$

\begin{equation}\label{eq:KV_extremal_rn}
\xi=\frac{\epsilon r_Nw^2}{2(r_Nw +m_e)^2}\partial_w -\left[c\frac{\epsilon r_N m_e(w\lambda)}{(r_N w + m_e)^3}+r_N\right]\partial_\lambda.
\end{equation}

The asymptotic norm of the Killing vector  gives a  timelike unit vector by construction. 
Observe the difference to the nonextremal case \eqref{eq:KV_nonextremal}, where the null surface $\lambda=0$ is not a Killing horizon for \eqref{eq:KV_extremal_rn}. 
In fact, the  boundary surface has an expansion rate proportional to $r_N$. 
An additional surface (to $w=0$) where the Killing vector becomes null is at
 \begin{equation}
 \lambda(w)=-\frac{2\epsilon(r_Nw+m_e)^2} {w}.
 \end{equation}
 
 {which follows from the calculation of \eqref{norm_gKV_}}. 
 This is the Killing horizon $r=m_e$.

Indeed, the parameter $r_N$ in \eqref{eq:rextrem} and \eqref{eq:vextrem} can be eliminated using the scaling relations $w\rightarrow w/r_N$ and $\lambda\rightarrow r_N\lambda$, so that the final solution for the extremal Reissner-Nordstr\"om black hole  depends only on the physical parameter $m_e$, the Bondi mass, so that
\begin{widetext}

\begin{subequations}\label{eq:sol_eRN_fin}
\begin{align}
    r 
    =& m_e + w +\epsilon\left[\frac{ w^2}{2(w +m_e)^2} \right]\lambda,
    \\
   V=& -\frac{4m_e \lambda^2}{(m_e+w)}
   \left[
   \frac{\epsilon \lambda   w^3 -(m_e-2  w)(m_e+   w)^3
   }{ \lambda^2 w^4
   +4\epsilon \lambda  w^2(m_e+w)^3 +4(m_e+w)^6 }
   \right],
   \\
  \xi&=\frac{\epsilon w^2}{2(w +m_e)^2}\partial_w -\left[\frac{\epsilon m_e(w\lambda)}{( w + m_e)^3}+1\right]\partial_\lambda.
    \label{eq:final_KV_only_m}
\end{align} 
\end{subequations}
\end{widetext}
This solution was also obtained by us in \cite{Gallo:2021jxt} using an alternative hierarchical system of equations; note, however, that the $w$ used here must be identified with the $-w$ used in that reference. 
In difference to the traditional  outgoing/ingoing Eddington-Finkelstein coordinates, the $\{w,\lambda,x^A\}$ is a regular coordinate system at both horizons.

\subsection{FJNW solution in null coordinates}\label{sec:FJNW_null}

The simplest Killing vector \eqref{eq:KV_general_sol} has $A=1$ and $B=0$. 
The Killing vector becomes $\xi=\partial_w$ and its additional requirement to be  asymptotically timelike restricts the asymptotic frame to be a Bondi frame, i.e. $H=1$ so that $r\sim \lambda$ for large values of $\lambda$.  
Hereafter, we consider a real and uncharged scalar field solution possessing this symmetry. We will find the FJNW solution \cite{Fisher:1948yn,Janis:1968zz,Wyman:1981bd,Virbhadra:1997ie,Ota:2021mub} in affine-null coordinates, which has not been derived explicitly  as a solution of a hierarchical system of the Einstein-scalar field equations in affine-null coordinates. 

Considering $\xi=\partial_w$ for an uncharged ($q=0$), real scalar field, the system of equations (\ref{hyp_ev}) becomes
\begin{align}
\label{eq:r-JNW}
& r_{,\lambda\lambda} +\frac{\kappa}{2} r \Phi_{,\lambda}^2 =0\;,\\
& Z_{,\lambda} = 0\;,\label{eq:Z_FJNW}\\
\label{eq:L-JNW}
& L_{,\lambda}=
      -\frac{\epsilon (\lambda + Z)\Phi_{,\lambda}}{r }\;, \\
\label{eq:V-JNW}
& V_{,\lambda\lambda} =
 -\frac{1}{\lambda}\left(\frac{\lambda^ 2 }{r^2}\right)_{,\lambda}+\frac{ 2 Z r_{,\lambda}}{r^3} - \frac{\kappa \epsilon}{r }\Phi_{,\lambda} L\;,
\end{align}
where we used that $\mathcal{L}\rightarrow L$   
becomes real.
The wave equation gives \eqref{eq:E}
\begin{equation}\label{eq:wave_static}
    (r^2 V \Phi_{,\lambda})_{,\lambda} = 0.
\end{equation}

In the absence of a scalar field, $\Phi=0$, the solution is trivial, yielding the Schwarzschild solution  $V = 1+b/r$ with  $r = \lambda $, where $b$ is a constant \cite{Madler:2023aqc}. 
Insertion of these Schwarzschild values into the wave equation \eqref{eq:wave_static} gives us after integration \begin{equation}
\Phi_{,\lambda} = \frac{a}{\lambda^2 + b\lambda}\;,
\end{equation}
whose subsequent integration yields
\begin{equation}\label{eq:ans_phi}
    \Phi = -\frac{a}{b} \ln\Bigl(1+\frac{b}{\lambda}\Bigr )
    = -\frac{a}{\lambda} + O(\lambda^{{-2}}),
\end{equation}
where we applied the  asymptotic condition, $\lim _{r\rightarrow \infty}\Phi=0$, for the scalar field. As it is also shown in \eqref{eq:ans_phi}, the  parameter $a$ is the scalar monopole. 
Although $\Phi$ in \eqref{eq:ans_phi} is a test scalar field, we can use this ansatz for the scalar field initial data to solve the complete Einstein-scalar field equations in a consistent way. 
In particular, note that if $r^2V$ has the same form for the complete system as in the  Schwarzschild case, the scalar field will have the same form as in the Schwarzschild case. The solution of \eqref{eq:r-JNW} gives us  
\begin{equation}\label{eq:raabb}
 r =c_1 \lambda\Bigl(1+\frac{b}{\lambda}\Bigr)^{\frac{1}{2}(1-\delta)}+c_2 \lambda\Bigl(1+\frac{b}{\lambda}\Bigr)^{\frac{1}{2}(1+\delta)}\;,
\end{equation}
with \begin{equation}
\delta=\sqrt{1-\frac{2\kappa a^2}{b^2}}.
\end{equation} 
The reality of  $\delta$ requires $2\kappa a^2<b^2$.
Note that if $a=0$,  the scalar field vanishes, and therefore the solution must reduce to the Schwarzschild metric. In that case \eqref{eq:raabb} reduces to $$r=(c_1+c_2)\lambda+c_2b.$$ 
For $a=0$, both $r$ and $\lambda$ serve as affine parameters. Thus, we can freely choose $r=\lambda$, which leads to $c_1=1$ and $c_2=0$. For this reason, in the following we have
\begin{equation}\label{eq:rsol_FJNW}
 r =\lambda\Bigl(1+\frac{b}{\lambda}\Bigr)^{\frac{1}{2}(1-\delta)}\;.
\end{equation}
The asymptotic limit of \eqref{eq:rsol_FJNW} gives
\begin{equation}
r= \lambda + \frac{b}{2}\left(1-\delta\right) + O(\lambda^{-1}),
\end{equation}
which from \eqref{eq:r_inf} implies 
\begin{eqnarray}
    H&=&1,\\
    R_\infty&=&\frac{b}{2}\left(1-\delta\right).
\end{eqnarray}
Next,  using the notation as in \eqref{eq:Zinfijn}, the integration of \eqref{eq:Z_FJNW} yields $Z(\lambda)=Z_\infty$=const,  which is related to the Bondi mass by \eqref{eq:mbondir-zinfy}, i.e.
\begin{equation}\label{eq:zinfiusar}
Z_\infty=R_\infty-2m_B=\frac{1}{2}\left[b(1-\delta)-4m_B\right].
\end{equation}

Then integration of \eqref{eq:L-JNW} yields
\begin{equation}
\begin{split}
L =& L_\infty
+ \frac{2a\epsilon}{( \delta-1)b}\left(1+\frac{  b}{\lambda}\right)^{-\frac{1-\delta}{2}} \\
&+\frac{2\epsilon a Z_\infty}{b(\delta+1)}\left(
\frac{1}{\lambda}
-\frac{2}{b(\delta - 1)}\right)   \left(1+\frac{  b}{\lambda}\right)^{-\frac{1-\delta}{2}}. 
\end{split}
\end{equation}

The integration constant  $L_\infty$ vanishes because of the Killing vector $\partial_w$ and considering \eqref{eq:LP1}.
Next, integration of \eqref{eq:V-JNW} gives us
\begin{equation}
    \begin{split}
        V =&  V_1\lambda+V_0
+\frac{2\lambda}{b(\delta-1)}\left(1+\frac{b}{\lambda}\right)^\delta 
\\&+ \frac{2Z_\infty\left[b(\delta-1)-2\lambda\right]}{b^2(\delta^2-1)}\left(1+\frac{b}{\lambda}\right)^\delta.     \end{split}
\end{equation}

Performing an asymptotic expansion of $V$ as in \eqref{eq:V_inf} 
gives
\begin{equation}
    \begin{split}
        V=&
        \left[V_1 + \frac{(2\delta + 2)b - 4Z_\infty}{b^2(\delta^2 - 1)}\right]\lambda 
        +\left[V_0 + \frac{2b\delta - 2Z_\infty}{(\delta-1)b}\right]
        \\&+O(\lambda^{-1}).
    \end{split}
\end{equation}
Taking into account \eqref{eq:Uinf}, \eqref{eq:Vinf} and the Killing symmetry  implies $V_\infty=1$ and  $U_\infty=0$, which gives us
\begin{align}
V_0=&\frac{2Z_\infty-(1+\delta )b  }{(\delta-1 )b},
\\
    V_1 =&\frac{2(1+\delta  )b - 4Z_\infty}{b^2(1-\delta^2 )}.
\end{align}
These relations allow us to write $V$ as
\begin{equation}\label{eq_Vfinaljnw}
\begin{split}
    V = &\frac{2\lambda\Bigl[b(\delta+1)-2Z_\infty\Bigr]}{b^2(\delta^2-1)} 
\left[\left(\frac{\lambda+b}{\lambda}\right)^\delta-1\right]\\
&+\frac{2Z_\infty }{b(\delta+1)}\left(\frac{\lambda+b}{\lambda}\right)^\delta
.
\end{split}
\end{equation}
Finally,  { from the asymptotic} expansion of \eqref{eq_Vfinaljnw}, which gives 
\begin{equation}
  V= 1+\frac{b\delta}{\lambda}+O(\lambda^{-2}),  
\end{equation}
we can connect the Bondi mass with $\delta$ and $b$, namely $\delta b=-2m_B$. Non-negativity of the Bondi mass requires $b\leq 0$. Hence, from \eqref{eq:zinfiusar} we can express $Z_\infty$ in terms of $b$ and $\delta$ as $$Z_\infty=\frac{b(\delta+1)}{2}.$$

By replacing these values into \eqref{eq_Vfinaljnw} we arrive to our final expression for $V$
\begin{equation}
    V=\left(1+\frac{b}{\lambda}\right)^\delta.
\end{equation}

The metric is the one of the classical FJNW solution \cite{Fisher:1948yn,Janis:1968zz,Wyman:1981bd,Virbhadra:1997ie,Ota:2021mub}
\begin{equation}
\begin{split}
    ds^2 =& -\Bigl(1+\frac{b}{\lambda}\Bigr)^\delta dw^2 +2\epsilon dwd\lambda 
\\
&+ \lambda^2 \Bigl(1+\frac{b}{\lambda}\Bigr)^{1-\delta}q_{AB}dx^Adx^B.
\end{split}
\end{equation}  
The corresponding nontrivial components of the contravariant metric are
\begin{equation}
    g^{w\lambda} = \epsilon\;\;,\;\;
    g^{\lambda\lambda} = \Bigl(1+\frac{b}{\lambda}\Bigr)^\delta\;\;,\;\;
    g^{AB} = 
    \lambda^{-2} \Bigl(1+\frac{b}{\lambda}\Bigr)^{\delta-1}q^{AB}.
\end{equation}

\section{Summary}\label{sec:summary}

We discussed the CIBVP in spherical symmetry  of the Einstein equations minimally coupled to a complex and charged massless scalar field. 
Charting spacetime with an affine-null metric at a family of null hypersurfaces, the field equations can be cast into a hierarchical system of hypersurface-evolution equations of which the evolution equation is a  first order transport equation for the complex initial data of this system. The initial data consist of an arbitrary  profile of the scalar field on an initial null hypersurface. 
The boundary values required to solve the hypersurface equations are determined from the type of boundary at which the family of null hypersurfaces is specified. 
Different CIBVPs were discussed; i) an asymptotic CIBVP for large values of the affine parameter, ii) a  CIBVP at the central geodesic of spherical symmetry, and iii) two realizations of double-null CIBVPs. 

The solution of the asymptotic CIBVP is represented by two functions, one for a redshift factor $H$ and another being the complex scalar news function $\mathcal{N}_{\mathcal{B}}$ (which is  the $w$ derivative of the scalar field's monopole) and in addition there are three scalar values at a given coordinate value $w=w_0$, representing the complex monopole of the scalar field, the initial Bondi mass and the initial value of the global charge of the system. The redshift factor indicates whether an event horizon forms, i.e. $H\rightarrow 0$,  or whether the solution approaches asymptotically Minkowskian  values, $H\rightarrow 1$. Given a continuous redshift function $H>0$ for subsequent values of $w$ within the region of integration, an asymptotic inertial coordinate system can be established in which  $H=1$ in that region.  Such coordinate frame is called a Bondi frame. 
Indeed, if $\epsilon=-1$ and the asymptotic observer is in a Bondi frame,  the $w$ coordinate corresponds to the retarded time $u$, while if $\epsilon=1$ the meaning of the $w$ coordinate is the advanced time $v$. Evaluation of the field equations in a Bondi frame significantly simplifies the solution of the field equation. 
In such Bondi frame, it is easy to see that the news function completely determines the future($\epsilon=-1$)/past($\epsilon=+1$) behavior of the scalar monopole, the Bondi mass and the global charge. 
This behavior is the spherically symmetric scalar field analog to the asymptotic CIBVP for electrovacuum space times (see e.g. \cite{1969RSPSA.310..221V,Bicak:1998vz}).

If the boundary degenerates to the central world line of spherical symmetry so that  the family of null hypersurfaces are a family of null cones whose vertices are traced by the world line, the regularity conditions at the vertices completely fix the boundary conditions to constant values along the world line. 
As a consequence, the local evolution of the scalar field along the world line is fully determined by the (first) derivative of the scalar field at the world line. 
As such, by specification of finite initial data  at the vertex of an initial null cone $w=w_0$ and requiring the local regularity conditions for the hypersurface variables for {other values of $w$ within the region of integration}, one can completely  determine the evolution of the scalar field and the resulting structure of the metric {in that region.}  Numerical simulations of the real analog of the presented hypersurface-evolution system at a world line was presented first in \cite{Crespo:2019mcv} and whose continuation \cite{Madler:2024kks} also showed the regular center can be maintained in spacetimes forming black holes.

The double-null CIBVP was considered for the two cases where  the Misner-Sharp mass of the null boundary surface is either trivial or nontrivial. 
In both cases, the only required initial data on given null hypersurface $w=w_0$ is the first $\lambda-$derivative of the scalar field on the initial data surface. 
At the common intersection of the initial data surface $w=w_0$ and the null boundary, a value for the scalar field must be specified. 
Moreover, an analog to the news function, referred to as the {\it boundary news function}, needs to be specified to be able to find the boundary values required to solve the hypersurface-evolution system. Like for the asymptotic  CIBVP, this boundary news function needs to be specified {\it a priori} and is completely independent from the initial data profile on $w_0$.

For the case, where the null boundary surface has vanishing Misner-Sharp mass, in fact all but two of these boundary values are zero. 
One nontrivial and constant boundary  value is the first $\lambda-$derivative of the areal distance that can be set to {unity without loss of generality}. 
The other one is the first derivative of $g^{\lambda\lambda}$ evaluated at the null boundary, but this field only needs to be specified as the initial value at the cross section between the initial data surface and the null boundary.
This boundary value is, in fact, the inaffinity of the generating null vectors of the boundary surface, where $w=w_0$. 
Its values for $w\neq w_0$ on the boundary are completely determined by  the boundary news function. 
To summarize, the double-null initial-boundary value problem with a null boundary of zero Misner-Sharp mass requires (apart from trivial and unity values) the specification of the scalar field value and inaffinity at the intersection of the initial data surface and the null boundary, the specification of a free news function on the boundary and the specification of the $\lambda-$derivative of the scalar field on the initial data surface. 

If the Misner-Sharp mass of the boundary is nonzero, the   values giving rise to the mass and  expansion rates  of the generators of the boundary and  initial  data surface are to be specified at the cross section of the two surfaces. 
In addition, a value for the charge and the scalar field ought to be specified at this intersection. 
These initial values are propagated along the null boundary after specifying the boundary news function. 
The set of four differential equations governing this propagation is a hierarchical set of (ordinary) differential equations along the generators of the null boundary. 
The evolution of the boundary values is  completely decoupled  from the evolution of the initial data via the hypersurface evolution equations and it can therefore be done stand alone and prior to any evolution of the initial data. 
This is in contrast to the timelike null  problem at the world line, where the evolution of the scalar field along the world line requires data (the $\lambda$-derivative) depending on the neighborhood of the world line.

The different CIBVP formulations can be used for numerically investigating  different physical scenaria. 
The world-line-null problem is suited for studies of charged black hole formation and critical phenomena based on regular and asymptotically flat initial data of a charged scalar field. 
Currently some of these studies are ongoing as an extension of the previous work employing a real scalar field \cite{Madler:2024kks}. This work used a coordinate compactification of the affine parameter together with some rescaling of the metric field to map null infinity to a finite value \cite{Madler:2025oiy}. 
Regarding the formation of charged black holes, it is natural to ask how close to unity the  charge-to-mass ratio of such black holes can be found  by evolving  charged scalar fields. Such simulations would test numerically the recent proof of the third law of thermodynamics \cite{Kehle:2022uvc}, which is still outstanding. 
A confirmation of the disproof of the third law requires that extremal charged black holes (for which the  charge-to-mass ratio is equal to one) form in finite (computational) time. 
But, previous investigations only reach values of  $\sim0.6$ for the charge-to-mass ratio \cite{Torres:2014fga}. 
The double-null CIBVP with a massive null boundary is well suited to investigate scattering \cite{Baake:2016oku} or 
perturbations of a spherically symmetric charged black hole \cite{Gelles:2025gxi} since the horizon of the black hole can be taken as a boundary surface. 
Given the recent success of using null geodesics stability analysis via Lyapynov exponents \cite{Ianniccari:2024ltb} to explain black hole formation, critical phenomena \cite{Choptuik:1992jv} or even black hole shadows \cite{Gallo:2024wju}, it may be interesting to explore how the presented CIBVP can be adopted to investigate such phenomena numerically. A formulation of a double-null CIBVP at a horizon may be an option for such study.

To better understand a spherically symmetric charged black hole in affine-null coordinates, we derived the subextremal and extremal Reissner-Nordstr\"om solution  using the developed affine-null metric double-null CIBVP. 
For that aim, we also looked at the general  solution of Killing equations. 
It turns out that there is an entire family of Killing vectors depending on two free functions depending on the $w$ coordinate. 
Indeed, a restriction of the Killing vectors to be asymptotically timelike, while using the general solution of the asymptotically flat initial boundary value problem, showed that the frame at large radii does not necessarily need to   be a Bondi frame, where the redshift factor is unity. 
This is, in fact, only the case if the Killing vector has the simple form $\partial_w$. For subextremal and extemal black holes, our results of Sec.~\ref{sec:RN_null} show that corresponding Killing vectors have much more elaborated functional dependence.  

Assuming the simplest Killing vector, we were also able to explicitly derive  the FJNW solution  based on a  guess of the  scalar field in a Schwarzschild background.  
It would be interesting to see if a
 charged version of the FJNW solution can be singled out within the presented formalism.

\begin{acknowledgements}
The work of R.G. is supported by ANID FONDECYT Regular No. 1220965. E.G. acknowledges financial support from CONICET and SeCyT-UNC.
T.M. thanks  J. Winicour, G. Khanna and S. Liebling for  discussions of aspects  on the subject. We also thank O. Baake for reading and commenting of the manuscript.   
\end{acknowledgements}

\section*{DATA AVAILABILITY}

No data were 
created 
or
analyzed
in 
this 
study.
\begin{appendix}
 
\end{appendix}

\bibliographystyle{apsrev4-1}      
\bibliography{bib_civbp_exact}   

\end{document}